\documentclass[12pt]{article}
\usepackage{latexsym}
\begin{document}


\def\integer{\rm integer}
\def\pr#1{\phi_{#1}}
\def\mod{\rm mod}
\def\Rb{ {\bf R}}
\def\B{ {\cal B}}
\def\L{ {\cal L}}
\def\V{ {\cal V}}

\def\S2{ {\bf S^2}}
\def\H2{ {\bf H^2}}
\def\R2{ {\bf R^2}}

\def\pix{ {\bar\pi}}



\begin{center}

\vskip 0.5 cm

{\large \bf  Hyperbolic Magnetic Billiards on Surfaces of Constant Curvature}

\vskip 1 cm

 Boris Gutkin 

\vskip 1 cm

{\em Department of Physics of Complex Systems\\
The Weizmann Institute of Science\\
Rehovot 76100\\
ISRAEL}\\{\small E-mail: fegutkin@vegas.weizmann.ac.il}
\vskip 2.0 cm

\end{center}

\begin{abstract}
We consider classical billiards on surfaces of constant curvature,
where the charged billiard ball is exposed to  a homogeneous, stationary magnetic field perpendicular to the surface.

We establish sufficient conditions for hyperbolicity of the billiard dynamics, and give lower estimation for the Lyapunov exponent. This extends our recent results for non-magnetic billiards on surfaces of constant curvature.
Using these conditions, we construct large classes of magnetic billiard
tables with positive Lyapunov exponents on the plane, on the sphere and on the hyperbolic plane.

\end{abstract}

\newpage

\section{ Introduction and the Statement of  Main Results}

\noindent In the present work we consider the billiards  on  two-dimensional surfaces $M$ of constant Gaussian curvature $r$ in the presence of a homogeneous  magnetic field of magnitude $\beta$, which is perpendicular to $M$. Inside the billiard domain $Q$ the pointlike  particle of unit charge and mass  moves at unit velocity  along curves of constant geodesic curvature $\beta$  and reflects elastically  at the boundary $\partial Q$. In the following, we will call these dynamical systems as {\it magnetic billiards} and $M$ as {\it magnetic surface} if $\beta\neq 0$.

Magnetic billiards on the plane have been considered in many works \cite{br}, \cite{k}, \cite{bk}, \cite{ta1} and on the hyperbolic plane in \cite{ta2}. The study of such billiards is strongly motivated by mesoscopic physics, where such billiard models are used as simplified version of the mesoscopic devices in the presence of magnetic fields. In the present paper we treat the magnetic billiards simultaneously on all surfaces of constant curvature (sphere, plane and hyperbolic plane).  For all values of $r$ and $\beta$ we establish  a common  criterion for hyperbolicity of the billiard dynamics, whose geometric realization depends only on the type of linearized dynamics (geometric optics) on $M$. This extends our recent results \cite{GGS} for non-magnetic surfaces of constant curvature. 

The  dynamics on $M$   depend crucially  both on the curvature of the surface and on the strength of magnetic field. Firstly, let us consider  the  case $\beta =0$.
 On the plane  the neighboring trajectories separate only linearly with time, so that the motion of the point mass between collisions with the boundary is neutral.  Exponential separation of billiard trajectories can only occur if the successive reflections from the boundaries introduce sufficient instability. On the hyperbolic plane  the negative curvature induces exponential divergence of geodesics. Thus, the boundary of hyperbolic billiard can be neutral (i. e.,  with zero curvature), and the  chaotic dynamics will be provided by the metric. On the sphere, in contrast, any  two geodesics   intersect twice at focal points. Thus, the boundary reflections  have to compensate for the focusing effect of the sphere,  in order to produce chaotic dynamics. We will call the mentioned above types of linearized dynamics  arising on the plane,  hyperbolic plane and  sphere as {\it parabolic }, {\it hyperbolic } and {\it elliptic } respectively.

In the presence of a constant magnetic field, an additional focusing effect appears. To simplify matters, let us discuss first the planar case when $\beta\neq 0$. Consider an infinitesimal beam of trajectories which  emerges from the same point  at the time $t=0$ (fig. 1a). Then, by elementary calculations (see e.g., \cite{k}) the curvature of the infinitesimal beam at the time $t=s$ is given by

\begin{equation} \chi(s)=-\beta\cot(\beta s).\end{equation}  
Consider now the reflection of  infinitesimal beams at the boundary. Let $m$ be the bouncing point, let $C$ be the osculating circle at $m$ and let $m'$ be the second point in which the particle trajectory intersects $C$.
Denote by $\chi_{-}$,  $\chi_{+}$  the curvatures of the infinitesimal beam immediately before and after reflection. Then, the  change of the curvature  under the reflection \cite{ta1}, \cite{k}  at the bouncing point $m$ (see fig. 1b) is given by  

\begin{equation} \chi_{+}-\chi_{-}=2\beta\cot(\beta d),\end{equation}
where $2d=\overline{ m m'}$ is the signed length of particle trajectory between $m$ and $m'$  (when the curvature of the boundary at $m$ is positive, $2d$ is simply the time which the particle spends after (before) reflection in the osculating circle $C$). It is  a simple observation,  that eqs. (1), (2) are  actually the same ones, which have appeared in \cite{GGS} for the non-magnetic billiards on  the sphere of radius $\beta^{-1}$, where the parameters $s$ and $d$ have the same geometric meaning. Thus, the geometric optics (or linearized dynamics) on the plane in the presence of a magnetic field  $\beta$ is equivalent to  the geometric optics on the non-magnetic sphere of  radius $\beta^{-1}$. More generally, we  demonstrate in the body of the paper  the equivalence of  geometric optics on the surfaces of constant Gaussian curvature $r$ in the presence of magnetic field $\beta$ to the geometric optics on the non-magnetic surfaces of constant Gaussian curvature $r_{eff}=r+\beta^2$. As a consequence, the type of linearized dynamics (elliptic, parabolic or hyperbolic)  depends only on the sign of  $r_{eff}$.   We will call the parameter $r_{eff}$ as {\it effective curvature} of the surface and will refer to the cases $r_{eff}=0$, $r_{eff}<0$, $r_{eff}>0$ as  Parabolic (P), Hyperbolic (H) and Elliptic (E) respectively.

Up to now,  the study of hyperbolic billiards has been mainly restricted to the Euclidean plane (see \cite{tb} for review).  See, however, \cite{ta2} for some results on chaotic billiards on the hyperbolic plane, and \cite{ve1}, \cite{ve2}, \cite{kss} for some results on   hyperbolic billiards on a general Riemannian surface. In our recent work  \cite{GGS} we have generalized  Wojtkowski's criterion of hyperbolicity for planar billiards \cite{wo2} to billiards on arbitrary surfaces of constant curvature.
 On  the basis of the equivalence of  geometric optics on magnetic and non-magnetic surfaces of constant curvature, we extend in the present paper the criterion of \cite{GGS} to the case of magnetic billiards.

The hyperbolicity criterion in \cite{GGS} can be formulated in terms of a special class of trajectories, which generalize two-periodic orbits.  
 Let $Q$ be a billiard table
on a surface of constant curvature.  The billiard  map $\phi :V\rightarrow
V$ acts
on the phase space $V$, which consists of pairs $v=(m,\theta)$.
Here $m$ is the position of the ball on the boundary $\partial Q$ of $Q$,
and $\theta$ is the angle between the outgoing velocity and
the tangent to $\partial Q$ at  $m$. The billiard  map
preserves a natural probability measure $\mu$ on $V$.
We denote the images of $v$  after
$n$ iterations by $(m_{n+1},\theta_{n+1}) =\phi^{n} (v)$.  The trajectory
$\phi^{n} (v)$
is a {\it generalized two-periodic trajectory } (g.t.p.t.) if the following
conditions are satisfied:\\
\\
\noindent 1. The incidence angle and the curvature of the boundary
$\kappa_{n}$
at the bouncing points have period 2:
$\theta_{2n}=\theta_{2}$, $\theta_{2n+1}=\theta_{1}$, $\kappa_{2n}=\kappa_2$,
$\kappa_{2n+1}=\kappa_1$;\\

\noindent  2. The length of trajectory between consecutive bouncing points
is constant: $s=\overline{ m_{n}m}_{n+1} $ (see fig. 2). \\

\noindent If $\theta_i = \pi/2$,  the g.t.p.t. is an usual two-periodic orbit.

Along a  g.t.p.t. the linearized map $D_{v}\phi$ is two-periodic, and the
stability of a  g.t.p.t. is determined by $D_{v}\phi^2$.
For each surface of constant curvature,
the stability type of a g.t.p.t.  is
completely determined by the triplet of parameters $(d_{1},d_{2},s)$, where
$2d_{1}$ (resp. $2d_{2}$) is the signed length of the chord generated by the
intersection of  the trajectory $m_{1} m_{2}$ with the  osculating circle
at $m_{1}$ (resp. $m_{2}$). We shall use the symbol
$T(d_{1},d_{2},s)$ for the g.t.p.t. with parameters $(d_{1},d_{2},s)$.

Let us consider g.t.p.t.s for planar, non-magnetic billiards in some detail. Here $s$ is the euclidean distance between consecutive bouncing
points,  and  $d_{i}= r_{i}\sin\theta_{i}$, $i=1,2$,  where  $r_{i}$ are
the radii of  curvature of the boundary $\partial Q$ at the respective points.
If the curvature of the boundary at the bouncing point is zero we take $r_{i}=-\infty$ as the radius of  curvature and $d_{i}=-\infty$ respectively.
By an elementary computation,  $T(d_{1},d_{2},s)$ is unstable if and only if

\begin{equation}
 s\in\cases{ [ d_{1},d_{2}] \cup  [ d_{1}+d_{2},\infty) &  if
$ d_{1},d_{2}\geq 0 $\cr
[0,\infty) &  if $d_{1},d_{2}\leq 0 $\cr
 [0, d_{1}+d_{2}]\cup [ d_{1},\infty) & if $d_{1}\geq 0,d_{2}\leq 0 $. \cr}
\end{equation}
Moreover, the trajectory is  hyperbolic (i. e., strictly unstable) if
$s$ is in the interior of the corresponding interval,
and the trajectory is parabolic if $s$ is a boundary point (in the limiting case $d_{1}=d_{2}=-\infty $ the trajectory is parabolic for any value of $s$).

We consider two classes of unstable g.t.p.t.s. The g.t.p.t. $T(d_{1},d_{2},s)$ is B-unstable if in eq. (3) $s$ belongs
to a ``big interval":

\begin{equation}
 s\in\cases{[ d_{1}+d_{2},\infty) &  if $ d_{1},d_{2}\geq 0 $\cr
[0,\infty) &  if $d_{1},d_{2}\leq 0 $\cr
[ d_{1},\infty) & if $d_{1}\geq 0,d_{2}\leq 0 $.\cr} 
\end{equation}
On the contrary, if $s$ belongs to a ``small interval", then
$T(d_{1},d_{2},s)$ is S-unstable:

\begin{equation}
 s\in\cases{[ d_{1},d_{2}] &  if $ d_{1},d_{2}\geq 0 $\cr
[ 0, d_{1}+d_{2}] & if $d_{1}\geq 0,d_{2}\leq 0 $.\cr} 
\end{equation}
Note that a small interval shrinks to a point when $|d_1|=|d_2|$.

 It has been demonstrated in   \cite{GGS}, that the notions of B-unstable and S-unstable g.t.p.t.s are generalized to arbitrary surfaces of constant curvature, where the analogs of (3,4,5) exist. The concept of g.t.p.t.s and the associated structures make sense for
 billiard on any surface immersed in a magnetic field. Since the stability properties of the trajectories depend only on the linearized dynamics (geometrical optics) of the system, one has essentially the same stability intervals (in terms of the parameters $(s,d_{1},d_{2})$) for g.t.p.t.s on  the surface of constant Gaussian curvature $r$ immersed in the magnetic field $\beta$ and for g.t.p.t.s on the non-magnetic surface of constant Gaussian curvature $r_{eff}=r+ \beta^2$. As a consequence, one can extend the notions of B-unstable and S-unstable g.t.p.t.s to  magnetic surfaces of constant curvature as well.  

Equipped with the mentioned above definitions, we  are ready now to formulate the main results of the present paper.
Let $Q$ be a  billiard table on a magnetic  surface of constant curvature, and let $\lambda(v) \geq 0$ be the Lyapunov exponent of the billiard. With any point $v=(m_1,\theta_1)\in V$ of the phase space
we associate a {\it formal} g.t.p.t. $T(v)$. Let
$\phi(v)=(m_{2},\theta_{2})$. We set $d_1=d(v)$, $d_2=d(\phi(v))$ and
$s=\overline{m_{1} m_{2}}$ for the length of particle trajectory between $m_{1}$ and $ m_{2}$. Then  $T(v)$ is determined by the triple $(d_{1},d_{2},s)$.
The formal g.t.p.t. $T(v)$ can be realized as an actual g.t.p.t.
$T(d_{1},d_{2},s)$
in an auxiliary billiard table $Q_{v}$, constructed from the
boundary $\partial Q$ around $m_{i}$ (see \cite{GGS}). Let $\phi_v$ be the map in $Q_{v}$ corresponding to the g.t.p.t.
$T(v)$, and let  $\bar{\lambda}(v)=\lim_{n\to\pm\infty}\frac{1}{n}\log||D\phi^{n}_v||\ge 0$
be the Lyapunov exponent of  $T(v)$. Then the sufficient condition for hyperbolic dynamics in $Q$ can be formulated as follows (see also Theorem 1 for the alternative formulation in the body of the paper).\\

\noindent  {\bf  Main Theorem. } {\it If for $\mu$ almost every point  $v\in V$,  $T(v)$ is  B-unstable, and for $\mu$ almost every point  $v\in V$,  $T(\phi^{n} (v))$ is strictly B-unstable for some $n$, then the billiard in $Q$ is hyperbolic ($\lambda(v)$ is positive $\mu$ almost everywhere).}\\

After  deriving the  conditions, which insure that    g.t.p.t.s are  B-unstable (analogs of (4)),
 this theorem turns to be a geometric criterion
for hyperbolicity of the billiard dynamics. In particular,  for planar non-magnetic billiards, the Main Theorem yields Wojtkowski's criterion for hyperbolic dynamics \cite{wo2}. Let $Q$ be a billiard table satisfying the
assumptions of the Main Theorem.  Following the approach of Wojtkowski's \cite{wo2},  the metric entropy  $h(Q)=\int_{V}\lambda(v)\,d\mu$ of the billiard can be actually estimated from below (Theorem 2 in the body of the paper):

\begin{equation} h(Q) \geq \int_{V}\bar{\lambda}(v)\,d\mu .\end{equation}

The  paper is organized as follows. In Section 2 we provide the necessary
preliminaries and establish the relationship between the geometric optics (i. e., the rules of propagation and reflection of infinitesimal light beams) on magnetic and non-magnetic surfaces of constant curvature.
In Section 3 we apply these results to obtain explicit analogs of
 (3-5) for all  magnetic (non-magnetic) surfaces of constant Gaussian curvature.
We obtain a linear instability conditions
for g.t.p.t.s  and show that they  distinguish  between   B-unstable and
S-unstable trajectories in a natural way. In Section 4, we reformulate the Main Theorem in a slightly different way and prove it 
using the invariant cone fields method.  We define our cone fields for magnetic billiards on all
surfaces of constant curvature exactly as in \cite{GGS}. Using the geometric optics language, the proof of the preservation of the cone field under the assumptions of the Main Theorem is reduced to the corresponding non-magnetic problem.
The Lyapunov exponent estimation (6) follows  from the results of \cite{GGS} by the same arguments. 
 In  Section 5, for each type of linearized dynamics   we derive  the criterion of hyperbolicity for elementary billiard
tables  (the boundary of these billiards consists of  arcs of constant geodesic curvature).
 We apply  it to construct several classes of magnetic  billiard tables with hyperbolic dynamics  on surfaces of constant curvature.
Finally, for each type of linearized dynamics we formulate some general principles  for the design of the magnetic billiard
tables satisfying the conditions of the Main Theorem. Several examples of billiards satisfying these principles are also given here. The derivation of the differential condition on the boundary of billiards satisfying the Main Theorem is given in the Appendix.


\section{ Geometric Optics}

Let $M$ be a surface of constant Gaussian curvature $r$. We will distinguish three cases:  $M=\R2$ {\it -plane} ($r=0$), $M=\S2$ {\it -sphere} ($r>0$) and $M=\H2$ {\it -hyperbolic plane } ($r<0$). Let us consider  the dynamical system arising on $M$ from the motion of particle of unit charge, mass and velocity in the presence of homogeneous magnetic field   perpendicular to the surface.  We will denote by $M(r,\beta )$ the corresponding surface $M$ in the presence of magnetic field of strength $\beta $. We can assume, without loss of generality, that  $\beta\geq 0$.  The particle trajectories on $M(r,\beta )$  are curves of constant geodesic curvature $\beta$ (circles, paracycles or hypercycles). We call $g^{s}$ the corresponding flow on $M(r,\beta )$. 

Let $Q$ be  a connected domain in $M(r,\beta )$, with a piecewise smooth boundary $\partial Q$. The billiard in $Q$ is the dynamical system arising from the motion of a point mass inside  $Q$ under the action of magnetic field $\beta$, with specular reflections at the boundary. The phase space $V$ of the billiard consists of unit tangent vectors, with origin points on $\partial Q$, pointing inside $Q$. The first return associated with $V$ is the billiard map, $\phi:V\rightarrow V$. We denote the probabilistic invariant measure on $V$ as $\mu$ (for its realization in the planar case see e.g.,  \cite{bk}). We will use the standard coordinates $(l,\theta)$ on $V$,
where $l$ is the arclength parameter on $\partial Q$ and
$\theta,\ 0\leq\theta\leq\pi$, is the angle between the vector and
$\partial Q$. 

 Let $D_v \phi : T_v V \to T_{\phi(v)} V$ denote the derivative of $\phi$. In what follows, we are interested in the action of $D\phi$ on the projectivization $B$ of the tangent space $TV$, see  \cite{wo1}, \cite{wo2}. The space $B=\cup B_v$, which is the set  of straight lines in the tangent planes $T_{v}V$, $v\in V$, can be conveniently represented using the language of  geometric optics. An oriented curve $\gamma\subset M$, of class $C^2$, defines
a ``light beam'', i.e., the family of particle trajectories orthogonal to $\gamma$. The trajectories  intersecting $\gamma$ infinitesimally close to a point,
$m\in\gamma$, form an ``infinitesimal beam'', which is completely determined
by the normal unit vector $\vec{n}\in T_{m}M$ to $\gamma$, and by the geodesic
curvature $\chi$ of $\gamma$ at  $m$. Let $\vec{n}=v\in V$ (the point $m$ belongs to the boundary of $Q$). We denote
by ${\cal B}(v,\chi)$ the  infinitesimal  beam determined by the pair $(v,\chi)$, $v\in V$, $\chi\in \Rb \cup\infty$.

On the other hand, each pair $(v,\chi)$ uniquely defines the line $b(v,\chi)\in B$, see e.g., \cite{wo2}. Here the vector $v$ defines the corresponding plane $T_{v}V$, and the curvature $\chi$ defines the direction of the line. Thus, infinitesimal beams yield a geometric representation of $B=\cup B_v$, where $\chi$  can be used as a projective coordinate on the space $B_v$, see \cite{wo2}, \cite{GGS}. As a result, one can study the action  of $D\phi$ on $B$ in terms of the action  of $\phi$ on the curvature of  infinitesimal beams $\B$.

Let $\rho_{m}:T_{m}M\rightarrow T_mM$, $m\in\partial Q$ be the linear reflection about the
tangent line to $\partial Q$. We will use the same letters, $\phi$, $\rho$, and $g$, for the differentials
of these mappings. Since the billiard map is the composition:
\begin{equation}
 \phi = \rho\circ g, 
\end{equation}   
it remains to compute the action of $g$ and $\rho$ on the curvature of infinitesimal beams. In other words, we need to know how the geodesic curvature of an  infinitesimal beam changes: a) along  free-flight trajectory on $M(r,\beta)$; b) under reflection. \\
 
\noindent{\bf a) Bouncless propagation.} Let us consider the change of the geodesic curvature $\chi (s)$ of an infinitesimal  beam  along a particle trajectory on $M(r,\beta)$. As it has been shown in \cite{ta2}, \cite{ta3} (see also \cite{k} for the planar case) the free-flight evolution of $\chi$ satisfies the Ricatti equation:

\begin{equation}
\dot{\chi}=r_{eff} +\chi^2 ,
\end{equation}
where $r_{eff}=r+\beta^2$ is the effective  curvature of $M(r,\beta)$. For convenience we introduce also  the parameter  $\xi=|r_{eff}|^{\frac{1}{2}}$ (then $\xi^{-1}$ is the radius of the non-magnetic surface, which has the same linearized dynamics as $M(r,\beta)$). Let $\chi=\chi(0)$ be the geodesic curvature of an infinitesimal   beam at the initial point, set $\chi'=\chi(s)=g^s\cdot\chi$ be the geodesic curvature after  the time $s$ of free-flight evolution. Using eq. (8) one can immediately obtain the relation between $\chi$ and $\chi'$. We will distinguish three cases:\\

{\it (E) -Elliptic $(r_{eff}>0)$.} $M$ is either $\S2$, or $\R2$, or $\H2$ with $\beta^2>|r|$ (strong field).
\begin{equation}
\chi'/\xi=-\cot(\xi s)+ \frac{\sin^{-2}(\xi s)}{\cot (\xi s)-\chi/\xi}; 
\end{equation}

{\it (P) -Parabolic $(r_{eff}=0)$.} $M$ is $\H2$ and $\beta^2=|r|$, or $M$ is $\R2$ and $\beta=0$.
\begin{equation}
\chi'=-s^{-1}+\frac{s^{-2}}{s^{-1}-\chi};
\end{equation}

{\it (H) -Hyperbolic $(r_{eff}<0)$.} $M$ is  $\H2$ and $\beta^2<|r|$ (weak field).
\begin{equation}
\chi'/\xi=-\coth(\xi s)+ \frac{\sinh^{-2}(\xi s)}{\coth (\xi s)-\chi/\xi}.  
\end{equation}
Note, for $M=\H2$ in the case (E) the particle trajectories are circles (trajectories of finite length), in the case (P) they are paracycles (trajectories of infinite length, which touch the boundary of Poincar$\acute{\rm{e}}$ disc) and in the case (H) they are hypercycles (trajectories  of infinite length, which have two points on the boundary of Poincar$\acute{\rm{e}}$ disc).\\

\noindent{\bf b) Reflection.} Let $\chi_{-}$, $v_{-}\in T_mM$ be the geodesic curvature and the direction of the infinitesimal beam just before the reflection at the point $m\in M$. We denote  $\chi_{+}=\rho\cdot\chi_{-}$, $v\equiv v_{+}= \rho \cdot v_{-} \in V$ to be the geodesic curvature and the direction of the infinitesimal beam immediately after reflection. Let $\kappa$ be the curvature of $\partial Q$
at $m$, and let $\theta$ be the angle between $v$ and the positive
tangent vector to
$\partial Q$ at $m$. Then, the extension of the well known formula  for planar billiards (see e.g., \cite{si}) to  magnetic billiards on arbitrary surfaces (see \cite{ta2}, \cite{ta3}) gives

\begin{equation}
 \chi_+=\chi_- +2 K(v), \qquad  \hbox{ where }\qquad  K(v)=\frac{\kappa+\beta\cos\theta}{\sin\theta}.
\end{equation}
Using classical formulas for surfaces of constant curvature
(\cite{vi}, see also \cite{ta1}, \cite{ta2}), it is possible to give a geometric interpretation of the function $K(\cdot)$.

Let $v\in V$, and let $l\in\partial Q$ be the origin point of $v$. Let
$C(l)\subset M$ be the osculating circle (resp. paracycle or hypercycle if $M=\H2$ and
$|\kappa(l)| \leq \xi  $)
of $\partial Q$. The free-flight particle trajectory, $G(v)$, corresponding to $v$ intersects $C(l)$ at $l$ and at another point  $l'$. In the cases (E) and (P) ($r_{eff}\geq 0$) let $d(v)$ be one half of the signed distance between
$l$ and $l'$, along $G(v)$. To eliminate the ambiguity  (when $K(v)=0) $, we choose the following intervals for $d(v)$:  $d(v)\in[-\infty,\infty)$ in the case (P) and $\xi d(v)\in[-\pi/2,\pi/2)$ in the case (E).
In the case (H) ($r_{eff}< 0$) we will use the following classification of points
of the phase space $V$, see fig. 3. We say that $v\in V$ is of type $A$ (resp. $B$) if
$|K(v)|\geq \xi$ (resp. $|K(v)| < \xi$). Let $V^A, V^B$ be the corresponding
subsets of $V$. Then $V=V^A\cup V^B$ is a partition. We denote by $d^A(v)$ ($d^B(v)$) one half of the signed distance between $l$ and $l'$ along $G(v)$ if $v\in V^A$ (resp. $v\in V^B$). To unite both cases we will use the notation:

\begin{eqnarray}
d(v) = \cases{
d^A(v)& if $v\in V^A$ \cr
d^B(v)+i\frac{\pi}{2}& if $v\in V^B$. \cr
}
\end{eqnarray}
Then we have
\begin{eqnarray}
 K^{-1}(v)=\cases{
d(v)& in the case (P)\cr
\xi^{-1}\tan(\xi d(v))& in the case (E) \cr
\xi^{-1}\tanh(\xi d(v))&  in the case (H). \cr
} 
\end{eqnarray}
As one can see from eqs. (9-14), the geometric optics (described in terms of parameters $d$, $s$) depend only on the value $r_{eff}$. This fact allows to study the linearized dynamics problems on $M(r,\beta)$ using the corresponding results for the non-magnetic surfaces of the constant Gaussian curvature $r_{eff}=r+\beta^2$. The corresponding transition $M(r,\beta)\rightarrow M(r_{eff},0)$ is schematically illustrated by fig. 4.

\section{ Stability of Generalized Two-Periodic Trajectories }

Let $Q$ be a billiard table  on  $M(r,\beta)$. For each $v\in V$ let $t(v)$ be the corresponding past semitrajectory in $Q$. Consider the curvature evolution of  an infinitesimal beam along $t(v)$.  Starting with  ${\cal B}(\phi^ {-k}\cdot v,\chi)$ for arbitrary $\chi$, we obtain after $k$ steps forward the infinitesimal beam  ${\cal B}(v,\chi^{(k)})$,  $\chi^{(k)}=\phi^ {k}\cdot\chi$. Eqs. (9-11) and (12) describe the action of the
billiard map on the curvature of infinitesimal beams. Assuming $k$ to be infinity, we obtain a formal continued fraction corresponding to the semitrajectory $t(v)$:

\begin{equation}
{
c(v)=\chi^{(\infty)}=
a_{0}+
{b_{0}\over\displaystyle a_{-1}+
   { \strut b_{-1} \over\displaystyle
a_{-2}\cdots }} }. 
\end{equation}  
 The coefficients of the continued fraction are determined by
$d_i=d(\phi^{i}\cdot v)$, and by the lengths $s_i$  of consecutive
billiard segments  as follows:

\[ \begin{array}{lll}
 (P) & a_{i}=-2s^{-1}_{i}+2d^{-1}_{i}, &  b_{i}=-s^{-2}_{i};\\
 (E) & a_{i}=-2\cot(\xi s_{i})+2\cot(\xi d_{i}), &  b_{i}=-\sin^{-2} (\xi s_{i});\\
 (H) & a_{i}=-2\coth(\xi s_{i})+2\coth(\xi d_{i}), &  b_{i}=-\sinh^{-2} (\xi s_{i}). \end{array} \]
\noindent  The continued fractions (15) determines the stability type of the trajectory: $t(v)$ is unstable if $c(v)$ is convergent (see e.g., \cite{si}). Since for a given sequence of $d_i$ and $s_i$,  $c(v)$ is completely determined by  $r_{eff}$,  one can reduce the problem of stability trajectories on $M(r,\beta)$ to the corresponding ``non-magnetic''  problem on $M(r_{eff},0)$.

As it has been mentioned in the introduction, we are  interested in the stability properties of generalized two periodic trajectories. 
A trajectory is a generalized two periodic trajectory (g.t.p.t.) if its parameters $d_i$ are periodic: $d_{2i+1}=d_1$, $d_{2i}=d_2$ and $s_i=s$ are the same along the trajectory (see fig. 2). Obviously, a g.t.p.t. yields a periodic continued fraction. We denote by  $T(d_1,d_2,s)$  the g.t.p.t. with parameters $(d_1,d_2,s)$ and by $c(d_1,d_2,s)$ the  associated continued fraction.

The stability of   $T(d_1,d_2,s)$, or equivalently, the  convergence of the two periodic continued fraction $c(d_1,d_2,s)$  has been studied in \cite{GGS} for non-magnetic surfaces of constant curvature. On the basis of the equivalence between the magnetic and non-magnetic  problems we can immediately generalize the results of \cite{GGS} to the  case $\beta\neq 0$. 
   
\vspace{2mm}
\noindent{\bf Proposition 1.} {\it The continued fraction $c(d_1,d_2,s)$
converges
if and only if the following inequalities are satisfied.}

 \[ \begin{array}{lc}
(P) & \qquad (s-d_{1})(s-d_{2})(s- d_{1}-d_{2})s\geq 0; \\
(E) & \qquad \sin(\xi (s-d_{1}))\sin (\xi(s-d_{2}))\sin(\xi(s-d_{1}-d_{2}))\sin (\xi s) \geq 0; \\
(H) & \qquad \sinh (\xi(s-d_{1}))\sinh (\xi(s-d_{2}))\sinh (\xi(s-d_{1}-d_{2}))\sinh(\xi s) \geq 0. \end{array} \]

Below we reformulate Proposition 1 explicitly as conditions for the instability  of the  corresponding g.t.p.t.

\noindent (P)   $T(d_1,d_2,s)$ is unstable if and only if

\begin{equation} s\in\cases{ [
d_{1},d_{2}] \cup  [ d_{1}+d_{2},\infty) &  if $ d_{1},d_{2}\geq 0
$\cr
[0,\infty) &  if $d_{1},d_{2}\leq 0 $\cr
 [0, d_{1}+d_{2}]\cup [
d_{1},\infty) & if $d_{1}\geq 0,d_{2}\leq 0 .$\cr} 
\end{equation}

\noindent (E) In this case $0 \geq \xi s \geq2\pi$, and we set $\pix=\pi\cdot\xi^{-1}$,

\[
 s\,\mod\pix = \cases{  s &  if $ s\le \pix $\cr
 s-\pix &  if $ s > \pix. $\cr}
\]

\noindent Then  $T(d_1,d_2,s)$ is unstable if and only if
\begin{equation} s\,
\mod\pix \in\cases{ [d_{1}+d_{2},\pix]\cup [d_{1},d_{2}] &  if $
d_{1},d_{2}\geq 0 $\cr
[0,d_{1}+d_{2}+\pix]\cup [\pix-d_{1},\pix-d_{2}] &  if
$d_{1},d_{2}\leq 0 $\cr
[d_{2},\pix+d_{1}]\cup [0,d_{1}+d_{2}] & if
$d_{1}\leq 0 , d_{2}\geq 0, |d_{2}|\geq|d_{1}|$\cr
[d_{2},\pix +d_{1}]\cup
[\pix+d_{2}+d_{1},\pix]   & if $d_{1}\leq 0 , d_{2}\geq 0, |d_{2}|\leq|d_{1}|.$\cr}
\end{equation}

\noindent (H) It matters whether $v_i\in V^A$ or $v_i\in V^B$
for $i=1,2$.
We say that  $T(d_1,d_2,s)$ is of type $(A-A)$ if $v_1\in V^A$ and $v_2\in
V^A$.
The other types: $(A-B)$, $(B-A)$, and $(B-B)$ are defined analogously.
We formulate the explicit criteria of instability for $T(d_1,d_2,s)$
type-by-type.

\noindent Type $(A-A)$:

\begin{equation}
s\in\cases{ [ d_{1}^{A},d_{2}^{A}] \cup  [ d_{1}^{A}+d_{2}^{A},\infty) &
if $ d_{1}^{A},d_{2}^{A}\geq 0 $\cr
[0,\infty) &  if
$d_{1}^{A},d_{2}^{A}\leq 0 $\cr
 [0, d_{1}^{A}+d_{2}^{A}]\cup [
d_{1}^{A},\infty) & if $d_{1}^{A}\geq 0,d_{2}^{A}\leq 0 $,\cr} 
\end{equation}
\noindent Type $(B-B)$:

\begin{equation}
 s\in\cases{ [ d_{1}^{B}+d_{2}^{B},\infty)&
if $d_{1}^{B}+d_{2}^{B} \geq 0 $\cr
[0,\infty) & if $d_{1}^{B}+d_{2}^{B}\leq 0
$,\cr}
\end{equation}
\noindent Types $(A-B)$ or $(B-A)$:

\begin{equation}
s\in\cases{ [
d_{1}^{A},\infty) &  if $d_{1}^{A}\geq 0 $\cr
[0,\infty) & if $d_{1}^{A}\leq
0$,\cr} 
\end{equation}
It is worth mentioning that in Proposition 1 (resp. eqs. (16-20))
the hyperbolicity of $T(d_1,d_2,s)$ corresponds to strict inequalities (resp.
inclusions in the interior). The equality case (resp. boundary case)
corresponds to the parabolicity of $T(d_1,d_2,s)$. There are also two special  cases when $T(d_1,d_2,s)$ is parabolic independently of the value of $s$: (P), $d_1=d_2=-\infty$ and (H), $|d_{1}^{A}|=|d_{2}^{A}|=\infty$. 

We call the right hand side of eqs. (16-20) the instability set of
$T(d_1,d_2,s)$.
In general, it is a union of two intervals, where
one of them  degenerates  when $|d_1|=|d_2|$,
while the other is always nontrivial.
Following the terminology of our previous work \cite{GGS}, we will say that the interval which persists is a
``big interval'', while the other one is a ``small interval''.  We will say that $T(d_1,d_2,s)$
is (strictly) B-unstable if $s$ belongs to the (interior of the) big
interval of instability.
The proposition below makes this terminology explicit.

\noindent{\bf Proposition 2.}  {\it The g.t.p.t. $T(d_1,d_2,s)$ is
B-unstable if (and
only if) the triple $(d_1,d_2,s)$ satisfies the following conditions:}

\noindent(P)
\begin{equation}
 s\in\cases{  [ d_{1}+d_{2},\infty) &  if $ d_{1},d_{2}\geq 0 $\cr
 [0,\infty) &  if $ d_{1},d_{2}\leq 0 $\cr
  [ d_{1},\infty) & if $d_{1}\geq 0,d_{2}\leq 0$ \cr}
\end{equation}
\noindent(E)
\begin{equation}
 s\,
\mod\pix\in\cases{ [d_{1}+d_{2},\pix] &  if $ d_{1},d_{2}\geq 0
$\cr
[0,d_{1}+d_{2}+\pix] &  if $d_{1},d_{2}\leq 0 $\cr
[d_{2},\pix+d_{1}]
& if $d_{1}\leq 0 , d_{2}\geq 0$\cr} 
\end{equation}
\noindent(H) The case $(A-A)$
\begin{equation} 
 s\in\cases{  [ d_{1}^{A}+d_{2}^{A},\infty) &  if $
d_{1}^{A},d_{2}^{A}\geq 0 $\cr
[0,\infty) &  if
$d_{1}^{A},d_{2}^{A}\leq 0 $\cr
 [ d_{1}^{A},\infty) & if $d_{1}^{A}\geq
0,d_{2}^{A}\leq 0, $\cr} 
\end{equation}

\noindent {\it or $|d_{1}^{A}|=|d_{2}^{A}|=\infty$ and arbitrary $s$.}

\noindent(H) The case $(B-B)$
\begin{equation} 
s\in\cases{ [ d_{1}^{B}+d_{2}^{B},\infty)&  if $d_{1}^{B}+d_{2}^{B} \geq 0
$\cr
[0,\infty) & if $d_{1}^{B}+d_{2}^{B}\leq 0 $\cr} 
\end{equation}
\noindent(H)  The cases $(A-B)$ or $(B-A)$
\begin{equation}
s\in\cases{ [ d_{1}^{A},\infty) &  if
$d_{1}^{A}\geq 0 $\cr
[0,\infty) & if $d_{1}^{A}\leq 0$.\cr}
\end{equation}
Obviously, the conditions (21-25) for B-unstable g.t.p.t.s are the same as those which appeared in \cite{GGS} for the corresponding non-magnetic cases.


\section{ The Main  Theorem}

Let $Q$ be a billiard table and $v\in V$ be an arbitrary point in the phase space of the billiard map.
Set $v_1=v, v_2=\phi(v), d_i=d(v_i), i=1,2$,
and let $s=s(v)$ be the length of the particle trajectory between the origin points of
$v_1$ and $v_2$  respectively. We will associate with $v$ a formal g.t.p.t. $T(v)=T(d_1,d_2,s)$, which parameters are defined by the triplet $(d_1,d_2,s)$.  We denote by $\lambda(v)$ the Lyapunov exponent of the billiard $Q$ and by $\bar{\lambda}(v)$ the  Lyapunov exponent of $T(v)$ (see Sect. 1), which are defined for $\mu$-almost all $v\in V$.

Using  Proposition 2 we introduce the following special class of points of the phase space of the billiard map.

\

\noindent {\bf Definition 1.} {\it A point $v\in V$ of the billiard phase space is

a)  B-hyperbolic (or strictly B-unstable) if the corresponding g.t.p.t. $T(v)$
 is strictly B-unstable;

b)  B-parabolic if the corresponding g.t.p.t. $T(v)$ is  B-unstable and
parabolic (i.e., $s$ belongs to the boundary  of the appropriate interval (21-25) );

c)  B-unstable if the corresponding g.t.p.t. $T(v)$ is  B-unstable
(i.e., B-parabolic or B-hyperbolic);

d)  eventually strictly B-unstable if there is some integer $n$ such that
 $T( \phi^{i} (v))$ is B-unstable for $0\leq i <n$ and $T( \phi^{n} (v))$ is
 strictly B-unstable. }

\

\noindent Below we  formulate the main theorem of the present work. 

\
\

\noindent {\bf  Theorem 1.} {\it Let $Q$ be a billiard table
on  $M(r,\beta)$. If    $\mu$-almost every 
 point of the billiard phase space is eventually strictly B-unstable, then the  Lyapunov  exponent  $\lambda$ is positive $\mu$-almost everywhere.}

\

\noindent {\it Proof.}  The proof of the theorem is based on the cone field method which has been initially applied to the planar billiards in \cite{wo1}, \cite{wo2}.

 A cone in $T_v V$ corresponds to an interval in the projectivization $B_v$. Therefore, a cone field, ${\cal W}$, is  determined by a function, $W(\cdot)$, on $V$, where each $W(v)$ is an  interval in the  projective coordinate $\chi$.   We define the function $W(v)$ as in  \cite{GGS}. For completeness, we repeat this definition below. 

\[ \begin{array}{cl}
   \hbox{ (P) and (E) }  & \qquad
 W(v)=\cases{ [K(v),+\infty]   &  if $ K(v)\geq 0 $\cr
  [-\infty,K(v)]  & if $K(v)\le 0$ \cr}\\ \\
  \hbox{ (H) }  & \qquad
 W(v)=\cases{ [K(v),+\infty]   &  if $ K(v)\geq \xi $\cr
  [-\infty,K(v)]  & if $K(v)\le \xi$ \cr}
 \end{array} \]

As it  follows from  Lemma 2 in  \cite{GGS}, this cone field is eventually strictly preserved by the billiard map if the conditions of  Theorem 1 are satisfied. By this fact the proof of the theorem  follows immediately  from   Wojtkowski's theorem (Theorem 1 in \cite{wo2}). \hfill$\Box$

Applying the method developed in \cite{wo2}, one can actually estimate from below the Lyapunov exponent using the cone field defined above. The result is given by the next theorem. 

\
 
\noindent {\bf  Theorem 2.} {\it Let $Q$ be a billiard table satisfying the assumptions of  Theorem 1, then 

$$h(\phi)=\int_{V}\lambda(v)\,d\mu \geq \int_{V}\bar{\lambda}(v)\,d\mu .$$
}

\

\noindent {\it Proof.} The proof follows immediately by the repetition of  calculations given in the proof of the analogous theorem for the non-magnetic case (see Theorem 2 in \cite{GGS}). \hfill$\Box$

\section{ Applications and Examples }   

 Theorem 1 together with Proposition 2  lead to a simple geometric criterion for billiard tables with hyperbolic dynamics. In this section we apply this criterion to construct various classes of hyperbolic billiards on $M(r,\beta)$.

\subsection{Elementary billiard tables} 
There is a class of billiard tables, where the application of  Theorem 1 gives an especially simple criterion for hyperbolicity. This class consists of billiard tables  $Q$, whose boundary is a finite union of arcs, $\Gamma_i$,  of constant geodesic curvature, $\kappa(\Gamma_i)=\kappa_i$. We call these tables elementary.  We will use the notation
$\Gamma_i^+$ (resp. $\Gamma_i^-$) if $\kappa(\Gamma_i)>0$
(resp. $\kappa(\Gamma_i)\leq 0$).
Let $C_i$ be the curve
of constant geodesic curvature such that $\Gamma_i\subseteq C_i$ and
 $D_i\subset M$ be the corresponding disk 
 ($C_i=\partial D_i$). Since the representation $\partial Q=\cup_{i=1}^N\Gamma_i$ is unique, we call $\Gamma_i$ the components of $\partial Q$. In the following, we consider elementary billiard tables for which  $|\kappa_i|\geq\beta$. One may easily see that the fulfillment of this inequality is necessary for billiards satisfying the conditions of Theorem 1 (see discussion in the Section 5.2 for billiards with boundaries of general type).

\

\noindent{\bf (E)} {\bf Elliptic case ($r_{eff}>0$).}
Let $D\subset M$ be a disc such that $\partial D$ is the circle whose geodesic curvature $\kappa$ satisfies $\kappa\geq \beta$. We define the component $-D\subset M$ as set of the points $m'\in M$ satisfying the condition $\overline{m'm}=\pix$  for some point $m\in D$, where $\overline{m'm}$ is the length of the particle trajectory between the points $m$, $m'$. We will refer to $-D$ as the dual component of $D\equiv +D$.  Straightforward analysis shows that $-D$ is the ring whose width equals to the diameter of $D$ and its radius is defined by $\xi$ (for $M=\R2$ its radius is $\beta^{-1}$), see figs. 5a,b,c. When $M=\S2$ and $\beta=0$,    $-D$ is the disk obtained from  $D$ by reflection about the center of $\S2$, as it has been defined in \cite{GGS}.  

Let us also introduce the  terminology:
If $R\subset S\subset M$ are regions with piecewise $C^1$ boundaries, we call an inclusion $R\subset S$ {\it proper} if $\partial R\cap
int\,S\ne\emptyset$.
The application of  Theorem 1 to the elementary billiard tables in the case $r_{eff}>0$ leads to the following criterion for hyperbolicity.

\

\noindent {\bf Corollary E.}\ {\it  Let $Q\subset M$ be an elementary
billiard table
whose boundary consists of $N>1$ components of type plus or minus.
Suppose $Q$ satisfies the following conditions:

 Condition E1. For every component
$\Gamma_{i}^{+}$ of $\partial Q$ we have $D_i\subset Q$. Besides, either
$-D_i\subset Q$, or $-D_i\subset M\setminus Q$, where the inclusions are
proper;

 Condition E2.  For every component
$\Gamma_{j}^{-}$ we have $D_j\subset M\setminus Q$, and the inclusions
$-D_j\subset M\setminus Q$, or $-D_j\subset Q$ are proper.

\noindent Then the billiard in $Q$ is hyperbolic.}

\noindent Outline of proof: The assumptions of Corollary E imply those of   Theorem 1.

\noindent{\it Remark}. Suppose $Q'= M\setminus Q$ is connected. If $Q$
satisfies Conditions E1 and E2, then $Q'$ also does, and hence the billiard
in $Q'$ is hyperbolic.

\

\noindent{\bf Examples}. 
{\it ``Lorenz gas'' billiards.} Such billiards are obtained by removing from $M$ a number of disjoint discs $D_i$, so that $Q=M\setminus\cup D_i$. If all the intersections $D_i\cap\pm D_j$ $i\neq j$, are empty, then the billiard in $Q$ is hyperbolic by Corollary E. The simplest example of such hyperbolic billiard is obtained by removing two disks from the magnetic plane, see fig. 6a.

The intersections $D_i\cap\pm D_j$ $i\neq j$, are always empty, if all the discs are contained inside of a free-flight particle trajectory (i.e., if all the discs lie  inside a circle of geodesic curvature $\beta$). Such billiards are the ``magnetic'' analogs of the non-magnetic hyperbolic billiard tables on the sphere, obtained by removing a finite number of disjoint disks from one hemisphere \cite{GGS}. The examples of hyperbolic billiards of this type on $\S2$, $\R2$ and $\H2$ are shown in fig. 6b,c,d. 

One can consider also unbounded   billiard tables $Q$ obtained by removing  an infinite number of disjoint disks from $\R2$, $\H2$. The simplest example of this type is obtained by removing a chain of equal disks from $M=\R2$, as shown in   fig. 7a  (this billiard can be also seen as cylinder with one hole). Because of the translation symmetry, one needs to check the non-intersection condition only for one disk. The  non-intersection  condition  is also necessary for hyperbolicity of such billiards. If it is not satisfied, then $Q$ has at least two stable g.t.p.t.s (see fig. 7a). 

Another type of unbounded hyperbolic billiard tables can be obtained by removing a lattice of the  disks from $M=\R2,\H2$. The example of such billiard shown in fig. 7b, is equivalent to the torus with one hole. Here, again, because of the translation symmetry, one has to check the non-intersection condition only for a single disk. 

{\it ``Flowers'' like billiards.} Consider a simply connected billiard table $Q$, whose boundary consists of several circular arcs of positive and  negative curvature satisfying the condition $|\kappa_i| \geq\beta$. Such billiards were originally introduced by Bunimovich \cite{bu1} \cite{bu2} as examples of planar (non-magnetic) hyperbolic billiards with convex boundary. It has been demonstrated  that for $r=0$, $\beta =0$ such billiards  are hyperbolic if the  conditions  $D_i\subseteq Q$ are satisfied for each convex component of the boundary. For $r_{eff}>0$  we have by Corollary E the additional requirement: $\partial Q\cap -D_i=\emptyset$ for each component of the boundary (compare with the analogous conditions in \cite{GGS} for the case $\beta =0$, $M=\S2$). The examples of hyperbolic billiards $Q$ on $\S2$, $\R2$, $\H2$ of ``flower'' type satisfying  the conditions of  Corollary E are shown in fig. 8a,b,c. It follows from the  remark above that  billiards in the domain $Q'=M\setminus Q$ are also hyperbolic.

\

\noindent{\bf (H+P)} {\bf Hyperbolic and parabolic cases ($r_{eff}\leq 0$).}
The criterion for hyperbolicity in this case is given by the following corollary.

\

\noindent {\bf Corollary H.} {\it Let $Q\subset M$ be an elementary
billiard table,
and let $\partial Q$ consist of $N>1$ components. If $Q$ satisfies conditions:

 Condition H1. For every convex component $\Gamma_i^{+}$ of $\partial Q$, we have $D_i\subset Q$;

 Condition H2.  For every concave component of $\partial Q$, we have $ \kappa(\Gamma_i^{-})\leq-\beta$ and for every convex component $ \kappa(\Gamma_i^{+})\geq\xi$.

\noindent Then the billiard in $Q$ is hyperbolic.}

\noindent Outline of proof: The assumptions of Corollary H imply those of the  Theorem 1.

\noindent{\it Remark}. When $\beta \to 0$ and $r\to 0$, the condition  H2 is automatically  fulfilled  and Corollary H turns to be the classical criterion of Bunimovich \cite{bu2} for hyperbolicity of planar, non-magnetic billiard tables.

\

\noindent{\bf Examples}. 
{\it Analogs of Sinai billiards.} The boundary of these billiards consists of concave arcs  $\Gamma_i^{-}$ of constant curvature (see fig. 9). If  the condition $\kappa(\Gamma_i^{-})\leq -\beta$ is satisfied for each component of the boundary, then the billiard is hyperbolic  by Corollary H.

{\it Analogs of Bunimovich billiards.} The example of hyperbolic billiard table with convex components satisfying the  conditions H1, H2 is shown in fig. 10.

\noindent{\it Remark}. The assumptions in Corollaries E and H that $N>1$
and that the inclusions be proper are needed only to exclude
certain degenerate situations, where each $v\in V$ is B-parabolic. This is
the case,
for instance, if $Q$ is a disc, or the annulus between concentric circles.

\subsection{ Hyperbolic billiard tables with boundary of general type}

Let us consider   billiard tables on $M(r,\beta)$ with piecewise smooth
boundary, $\partial Q=\cup_{i}\gamma_i$ of general type. The components $\gamma_{i}$ are $C^2$ smooth curves parameterized by the arclength $l$, whose curvature $\kappa_{i}(l)$ has the same sign along each $\gamma_{i}$. We will refer to $\gamma_{i}$ as convex component if $\kappa _{i}(l)>0$, or as concave component if $ \kappa_{i}(l)\leq 0$. Let us denote $\kappa( \gamma_{i})=\max \lbrace \kappa_i(l), l\in \gamma_{i}\rbrace $ for the convex components, and  $ \kappa(\gamma_{i})=\min \lbrace \kappa_i(l), l\in \gamma_{i}\rbrace$ for the  concave components. 

Following the terminology  in \cite{wo2}, we introduce the class of convex scattering curves on $M(r,\beta)$.

\noindent{\bf Definition 2.} {\it A smooth convex curve  $\gamma\subset M$ is (strictly) convex scattering if for any  $v\in V$,
such that the origin points of $v$ and $\phi(v)$ belong to $\gamma$, the corresponding g.t.p.t. $T(v)$ is (strictly) B-unstable. } \\

A curve $\gamma$ is convex scattering if  one of the relevant conditions (21-25) is satisfied for each pair of  points on $\gamma$. Regarding the planar non-magnetic case, this leads to the definition of Wojtkowski \cite{wo2} for convex scattering curve. Let us introduce the parameter $R(l)=(\kappa(l)-\beta)^{-1}$. Considering the infinitesimally close points on  $\gamma$ we show in Appendix that  the condition $R''(l)\leq 0$ is necessary for  $\gamma$ to be convex scattering. It should be noted, that this condition is also sufficient in the planar, non-magnetic case (see \cite{wo2}), but not for generic parameters $r$, $\beta$ (see \cite{GGS} for $\beta=0$ case).

In what follows, we formulate  the principles for design of hyperbolic billiards satisfying the conditions of  Theorem 1. Let $Q$ be a billiard table satisfying the conditions of Theorem 1. Then each convex component of $\partial Q$ has to be  convex scattering and consequently, the condition $R''\leq 0$ holds along each convex component of the boundary. 
There is an additional  restriction on the curves $\gamma_i$ which compose the boundary of $Q$. It  follows from Proposition 2 that for billiards satisfying the conditions of  Theorem 1 the sign of $K(v)$ ($d(v)$)  depends only on the origin point of $v$ (there is no dependence on $\theta$) for any $v\in V$, i.e.,   $K(v)$ ($d(v)$) has the same sign along $\gamma_i$ as $ \kappa(\gamma_{i})$. This happens if    for each component  $\gamma_{i}$, $|\kappa( \gamma_{i})|\geq\beta$ (the magnetic field is sufficiently weak). Thus, in what follows we particularly  exclude from our consideration the magnetic billiards with flat boundaries. Such  billiards do not satisfy the conditions of  Theorem 1.

\vspace{3mm}
\noindent  {\bf Design of hyperbolic billiard tables in the  (E) case}.

\noindent By Definition 2 a curve $\gamma$ is convex scattering if it is convex and the condition 

\begin{equation}
d_1+d_2\leq s \leq \pix,
\end{equation}
holds for any pair of points on $\gamma$. For simplicity of exposition, we will restrict our attention for $M=\R2, \H2$ to the bounded billiard tables and for $M=\S2$ to the billiard tables which can be placed in a hemisphere. Theorem 1 yields the following principles for the design
of piecewise  billiard tables with hyperbolic dynamics in  (E) case:

\noindent P1: $|\kappa(\gamma_i)|\geq\beta$ for all components. 

\noindent P2: All convex components of $\partial Q$ are convex scattering.

\noindent P3: Any convex component of $\partial Q$ has to be ``sufficiently far'', but not ``too far'', from any other component. Any concave component has to be not ``too far'', from any other concave component.

The precise meaning of  P3 is that the parameters of any two consecutive bouncing points, which belong to  different components of the boundary, satisfy the condition (22). In particular it implies the set of restrictions on the angles between consecutive components of the boundary. It can be formulated as an additional principle. 

\noindent P4:  Let $\gamma_{i},\gamma_{i+1}\subset\partial Q$ be two adjacent components,
meeting at a vertex. If both $\gamma_{i}$ and $\gamma_{i+1}$ are convex, then
the interior angle at the vertex is greater than $\pi$.
If $\gamma_{i}$ and $\gamma_{i+1}$ have different sign of curvature, then the angle in question is greater  or equal to $\pi$.

Another restriction which arises from P3 is that  the length (equivalently  the time) of free-flight between any two  consequent bouncing points on the boundary of the billiard has to be not greater than $\pix$. In other words,   the  billiard table has to be  ``smaller'' than  circle drawn by a free-flight particle on $M(r,\beta)$.

\

\noindent{\bf Examples.}
The examples of the hyperbolic billiards on $\R2$ satisfying the above principles are shown in fig. 11a,b. A bounded Sinai-like billiard, whose boundary consists of (strictly) concave components (fig. 11a) always satisfies the principles P1-P4 for sufficiently weak magnetic field. 

The example of a convex billiard is shown in fig. 11b. It is a cardioid, whose boundary is  strictly convex scattering curve for $\beta=0$ (see \cite{wo2}). For $\beta=0$ this billiard is hyperbolic, as it follows from Theorem 1. Since  strictly convex scattering curve remains to be such under small perturbations of $\beta$, the  billiard in fig. 11b is hyperbolic for sufficiently weak magnetic field.

\

\noindent  {\bf Design of hyperbolic billiard tables in the  (P+H) case}.

\noindent Definition 2 leads  to the following geometric conditions on the convex scattering curve in the (P+H) case. A convex curve $\gamma$ is convex scattering if $\kappa(\gamma)\geq \xi$ and for each pair  of points on $\gamma$
\begin{equation}
d_1+d_2\leq s. 
\end{equation}

Theorem 1 yields the following principles for the design
of billiard tables with hyperbolic dynamics in the (P+H) case:

\noindent P1: $\kappa (\gamma_i)\geq \xi$ for any convex component of $\partial Q$ and $\kappa (\gamma_i)\leq -\beta$ for any concave component of $\partial Q$. 

\noindent P2: All convex components of $\partial Q$ are convex scattering.

\noindent P3: Any convex component of $\partial Q$ is ``sufficiently far'' from
any other component.

More precisely, condition P3 means that any two
consecutive bouncing points of the billiard ball, which belong to different components,  satisfy eqs. (23-25). In particular, this
yields, the same inequalities (P4) as in  (E) case, for the interior angles between consecutive components of
$\partial Q$.

\

\noindent{\bf Examples.}
In  (P+H) case, any concave billiard is hyperbolic if the condition $\kappa(\gamma_i)\leq -\beta$ is fulfilled for each component of the boundary. As in the  case (E), the examples of the convex hyperbolic billiards can be obtained from their non-magnetic counterparts satisfying the conditions of Theorem 1.

Finally, it should be noted, that formulated above principles for design of hyperbolic billiards on $M(r,\beta)$ are robust under small perturbations of $\beta$, $r$ and the billiard wall. Generally, one can construct hyperbolic  billiards on magnetic surfaces of constant curvature on the basis of the corresponding non-magnetic planar billiards satisfying Wojtkowski's criterion. 


\section { Conclusions}
In the present paper we have formulated the  criterion for hyperbolic dynamics in billiards on surfaces of constant Gaussian curvature $r$ in the presence of a homogeneous magnetic field $\beta$ perpendicular to the surface. The criterion is valid for all values of $r$, $\beta$   and its geometric realization   depends only on the type of linearized dynamics (elliptic, parabolic or hyperbolic). In this way we extend our recent results in \cite{GGS} to the case of magnetic surfaces of constant curvature. The basic property, which allows  unification of the hyperbolicity criteria for the magnetic and non-magnetic billiards on surfaces of constant curvature, is the equivalence between the geometric optics in both cases. In fact, in terms of special parameters $d_i$, $s_i$ the  geometric optics depend only on the effective curvature $r_{eff}=r+\beta^2$ of the surface.  It is important to stress, that the dynamics in  magnetic and non-magnetic billiards are very different (e.g., the magnetic field breaks time reversal symmetry). It is the only  linearized dynamics, which are the same for the considered systems. Applying the hyperbolicity criterion, we were able to construct the different classes of hyperbolic billiards for each type of the linearized dynamics (equivalently for each of the signs of $r_{eff}$).

There are two types of necessary conditions which arise for hyperbolic billiards satisfying
our criterion. The first one is a requirement  for the  convex components of the boundary to be convex scattering. As a consequence, the inequality $R''(l)\leq 0$ has to be satisfied along each convex component. This inequality is generalization of well-known Wojtkowski's condition \cite{wo2} for convex component of planar (non-magnetic) hyperbolic billiard.  It has been demonstrated for planar non-magnetic billiards in \cite{bu3}, \cite{bu4}, \cite{do} that Wojtkowski's criterion can be considerably strengthened. This suggests, in particular, that  condition $R''(l)\leq 0$ can be relaxed for general parameters $r$, $\beta$ by employing  invariant cone fields, different from the  one used in the present paper (see  discussion in \cite{GGS}).  
The second type of conditions is specific for magnetic billiards. This is a  requirement of ``weakness'' for  the magnetic field compared to the curvature of the billiard boundary.  For generic systems, such condition is expected, in order  to prevent stable skipping orbits close to the boundary. It has been shown in \cite{br}, (see also \cite{bk}) that billiard with sufficiently smooth boundary possesses invariant tori corresponding to skipping trajectories. It seems that in the strong field regime a part of stable   periodic orbits has to survive even if the smoothness of the boundary is broken. It remains, however  an open question, whether the condition  $|\kappa_i|\geq \beta$ can be relaxed for generic billiard.

The positive Lyapunov exponent for a billiard   implies strong mixing properties: countable number of ergodic components, positive entropy, Bernoulli property etc. It should be pointed out, however, that ergodicity does not automatically  follow from the positivety of Lyapunov exponent. Nevertheless, 
one can expect that  billiards satisfying the conditions of Theorem 1  will be  typically ergodic. It seems that the methods  developed for the proof of ergodicity of planar hyperbolic  billiards can be extended to the class of billiards considered in the present paper.


\section* { Acknowledgment}

The author is indebted to Professor U. Smilansky for proposing this investigation and for critically reading the manuscript. The author would like to thank Andrey Shapiro de Brosh for interesting and inspiring discussions, and various valuable remarks.

 This work was supported  by the Minerva Center for Nonlinear Physics of Complex Systems.


\section { Appendix }

We will investigate the conditions under which a convex arc on the surface of constant curvature $M$ 
in the presence of magnetic field $\beta$ is convex scattering.

For simplicity of exposition, we consider the case, when $M$ is magnetic plane.  Let $\gamma(l)\subset M$  be any smooth curve parameterized by arclength $l$, and let $\kappa(l)$ be the geodesic curvature of $\gamma$. 
Let  now $\gamma (l_{0})$ and $\gamma (l_{1})$ be two points on
$\gamma$, such that the arc of  $\gamma$ between $\gamma (l_{0})$ and
$\gamma (l_{1})$ lies entirely on one side of straight line passing through $\gamma (l_{0})$ and $\gamma (l_{1})$. We choose cartezian coordinate  system $(x,y)$ in such a way that $y(l_0)=y(l_1)=0$, $x(l_0)=-x(l_1)$ and the arc of $\gamma$ between $\gamma (l_{0})$ and $\gamma (l_{1})$ lies above  $x$-axis, see fig. 12. 
  Let $\alpha(l)$ be the angle, which $\frac{d\gamma}{dl}$ makes with $x$-axis , then  

\begin{equation}
\frac{dx}{dl}=\cos\alpha ; \qquad \frac{dy}{dl}=\sin\alpha ; \qquad
\frac{d\alpha}{dl}=-\kappa.
\end{equation}
We introduce also an auxiliary variable $\delta$, such that $\beta x=\sin\delta$. 

 For $\beta>0$ there are two different particle trajectories connecting the points $\gamma (l_{0})$ and $\gamma (l_{1})$ (resp. two different g.t.p.t.s corresponding to these points), see fig. 12. Below, we consider the trajectory which lies in the lower halfplane. Then, the results for  trajectory in the upper halfplane are obtained by the change of the sign of $\beta$ to the opposite. Let
 $\theta=\alpha+\delta$.  Then,  at the points $l_{0,1}$,  $\theta(l_{0,1})$  are the angles between $\gamma$ and the particle trajectory connecting $\gamma (l_{0})$ and $\gamma (l_{1})$.
Set $\Delta =s-d_{1}-d_{2}$. By eq. 14 we get

\begin{eqnarray*} 
\Delta = \beta^{-1}\int\left[ d\left(\arctan\left(\frac{\beta\sin\theta}{\kappa-\beta\cos\theta}\right)\right) +d\delta \right]
\end{eqnarray*}

\begin{equation} 
 =\int dl \left(\frac{-\kappa'\sin\theta+\kappa\left( \kappa+\beta\frac{\sin\alpha}{\sin\delta} \right)\left( \frac{\cos\alpha}{\cos\delta}- \cos\theta\right)}{\kappa^2+\beta^2 -2\beta \kappa\cos\theta}\right)  
\end{equation}
We separate the last integral into the sum of  two parts.
The first one is

\begin{eqnarray*}
I= \int dl \left(\frac{-\kappa'\sin\theta}{\kappa^2+\beta^2 -2\beta \kappa\cos\theta}\right)=
\int dy \left(\frac{ R'}{1+4R^2 \kappa\beta\sin^2\theta/2} \right),
\end{eqnarray*}
where  $R^{-1}(l,\beta)=\kappa(l)-\beta$. Since $y(l_0)=y(l_1)=0$, we obtain 

\begin{equation}
I=-\int dl\left(\frac{ y R''}{1+4R^2 \kappa\beta\sin^2\frac{\theta}{2} } -
\frac{y R'(4R^2 \kappa\beta\sin^2\frac{\theta}{2})'}{(1+4R^2 \kappa\beta\sin^2\frac{\theta}{2})^2}\right)=
-\frac{R''L^3 \kappa}{12} + O(L^4),
\end{equation}
where $L=l_1 -l_0$ is the length of the curve between the points $\gamma (l_{0})$, $\gamma (l_{1})$. Analogously, for second part we have

\begin{eqnarray*}
II=\int dl \left(\frac{\kappa\left( \kappa+\beta\frac{\sin\alpha}{\sin\delta} \right)\left( \frac{\cos\alpha}{\cos\delta}- \cos\theta\right)}{\kappa^2+\beta^2 -2\beta \kappa\cos\theta} \right)
\end{eqnarray*}

\begin{eqnarray*}
= \int dy \left(\frac{\kappa\frac{\sin\theta}{\cos\delta}\left( \kappa\frac{\sin\delta}{\sin\alpha}  +\beta \right)}{\kappa^2+\beta^2 -2\beta \kappa\cos\theta} \right)     =O(L^4).
\end{eqnarray*}
 Adding both parts we obtain finally

\begin{equation}
\Delta = I+II =-\frac{R''L^3 \kappa}{12} + O(L^4).
\end{equation}
Thus, if the curve $\gamma$ is convex scattering, then the condition $R''(l,\beta)\leq 0$ holds everywhere on $\gamma$. Considering trajectories of the second type (i.e., trajectories which lay in the upper halfplane), we obtain the condition
$R''(l,-\beta)\leq 0$ for convex scattering curves. However, it easy to see, that 
$R''(l,\beta)\leq 0$ actually implies  $R''(l,-\beta)\leq 0$.

Repeating the same analysis for general $M(r,\beta)$ we have found (see also \cite{GGS} for $\beta=0$ case)  that eq. 31 holds 
for all surfaces of constant curvature. As a consequence,  $R''\leq 0$ is a necessary condition for convex scattering on $M(r,\beta)$. On the contrary, if the strict inequality  $R''< 0$ holds along $\gamma$, then by eq. 31, any sufficiently small piece of $ \gamma$ is convex scattering. 


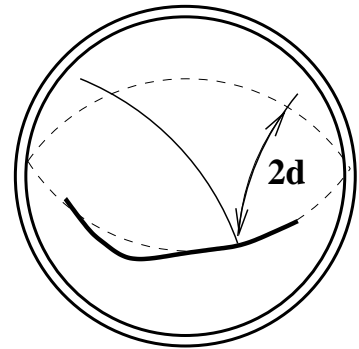
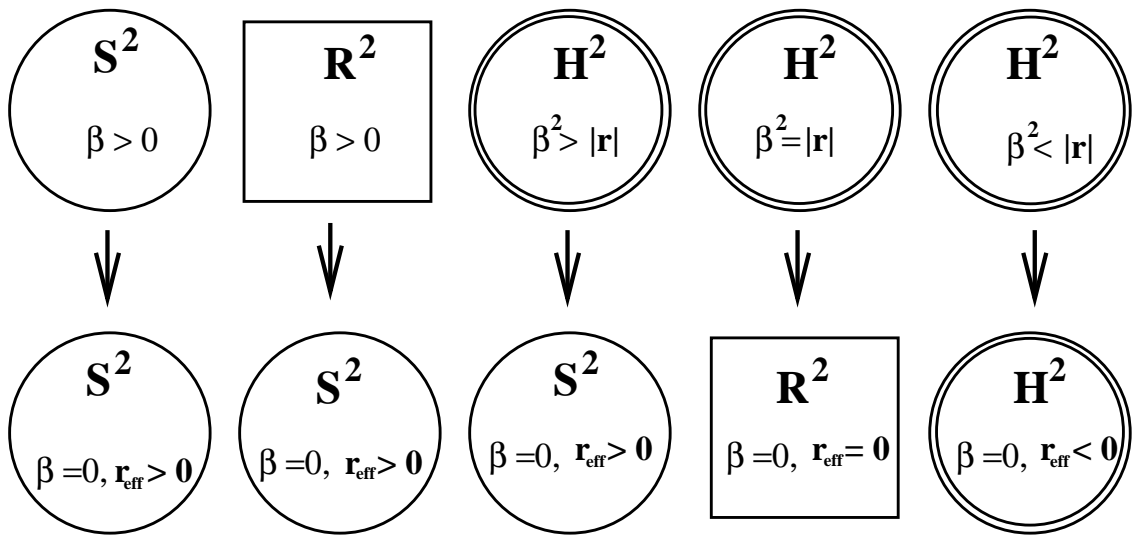
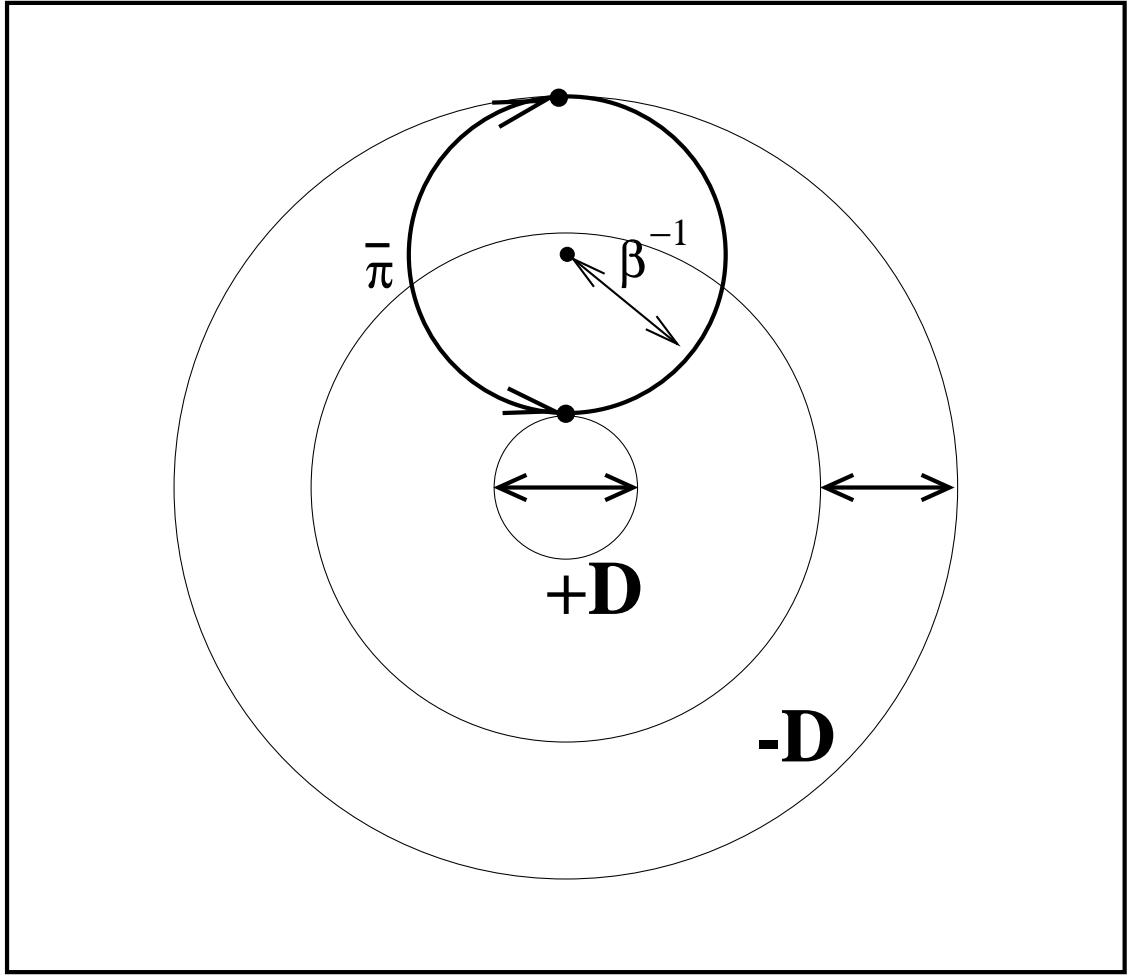
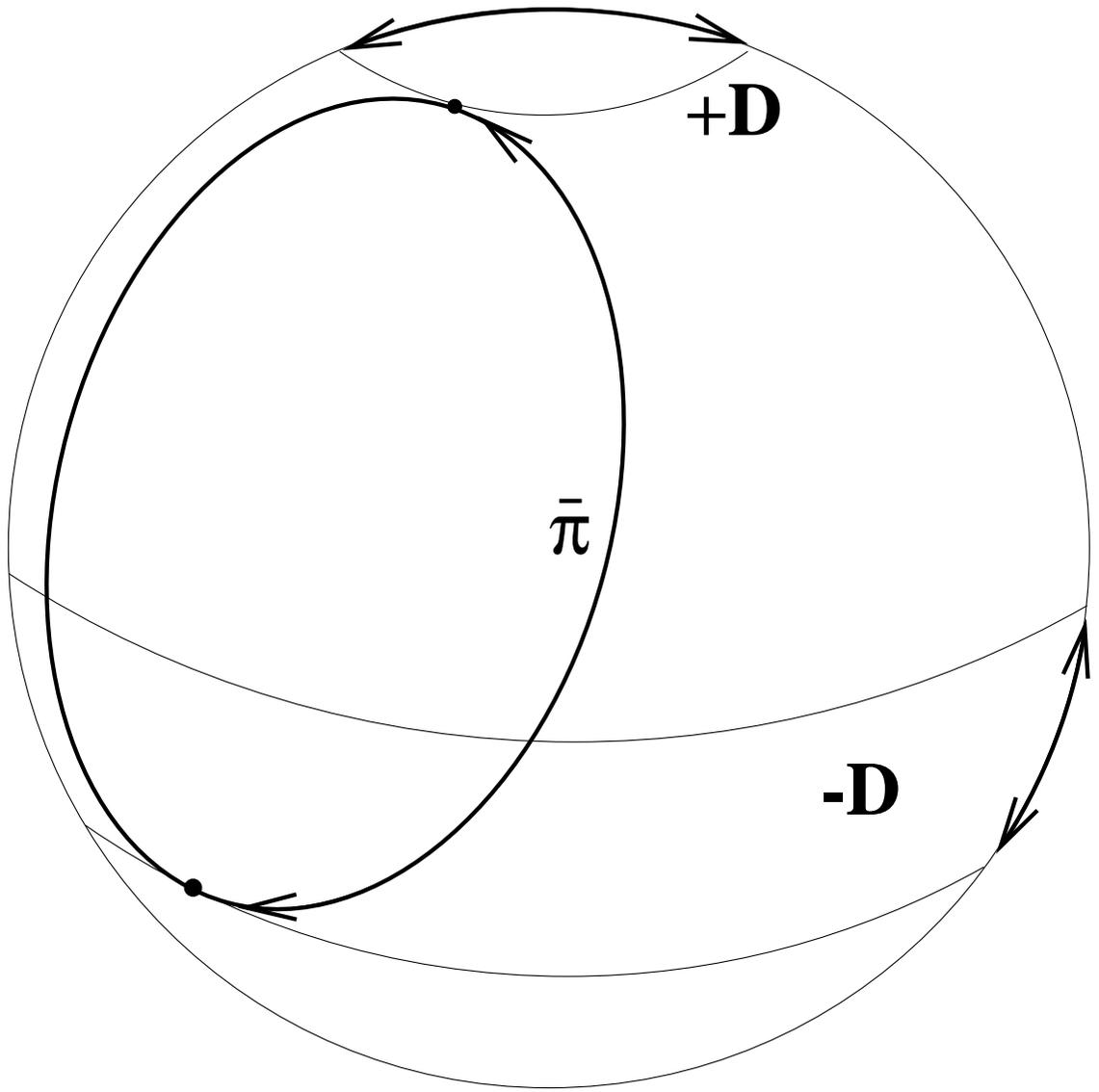
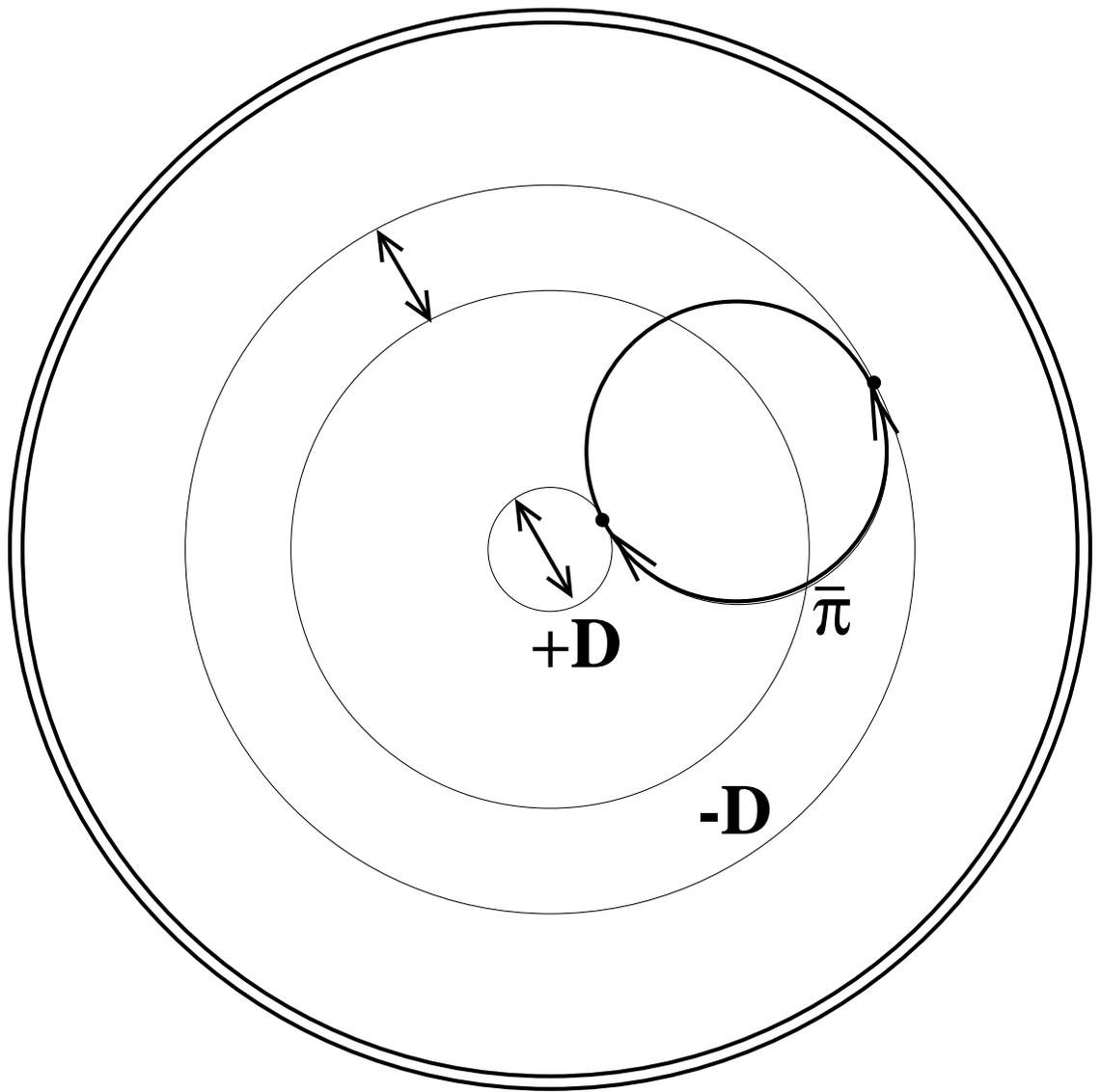
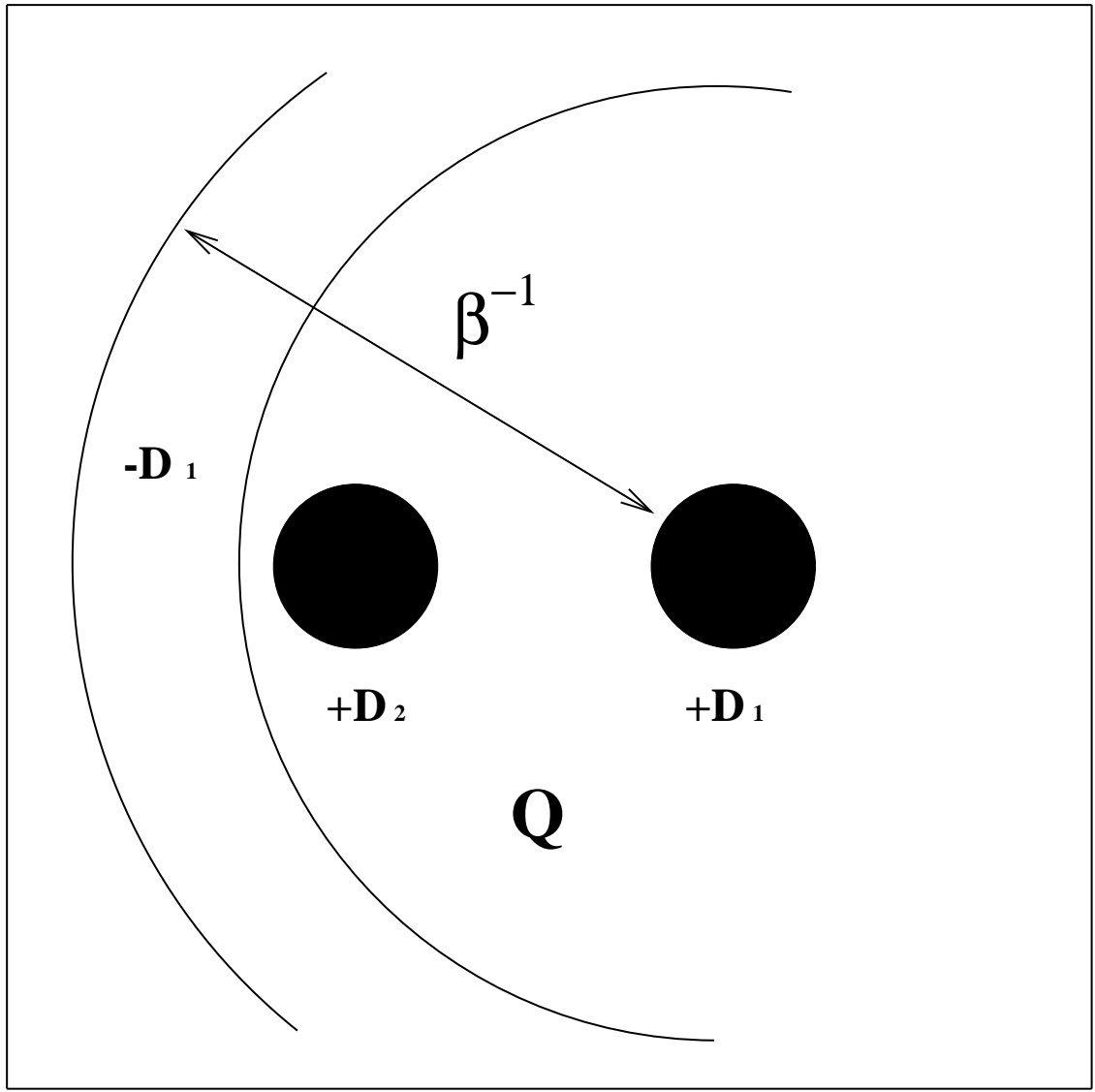
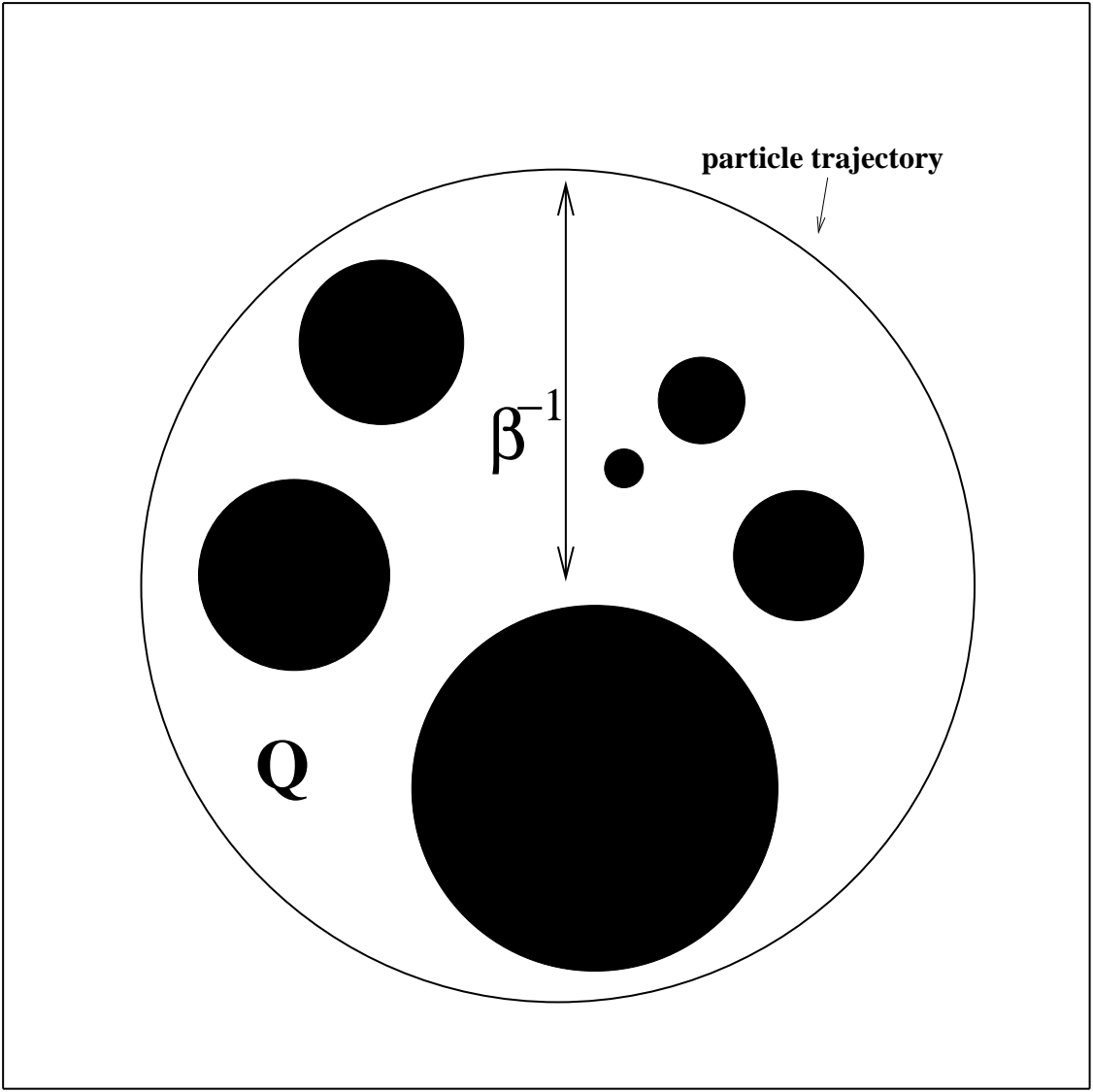
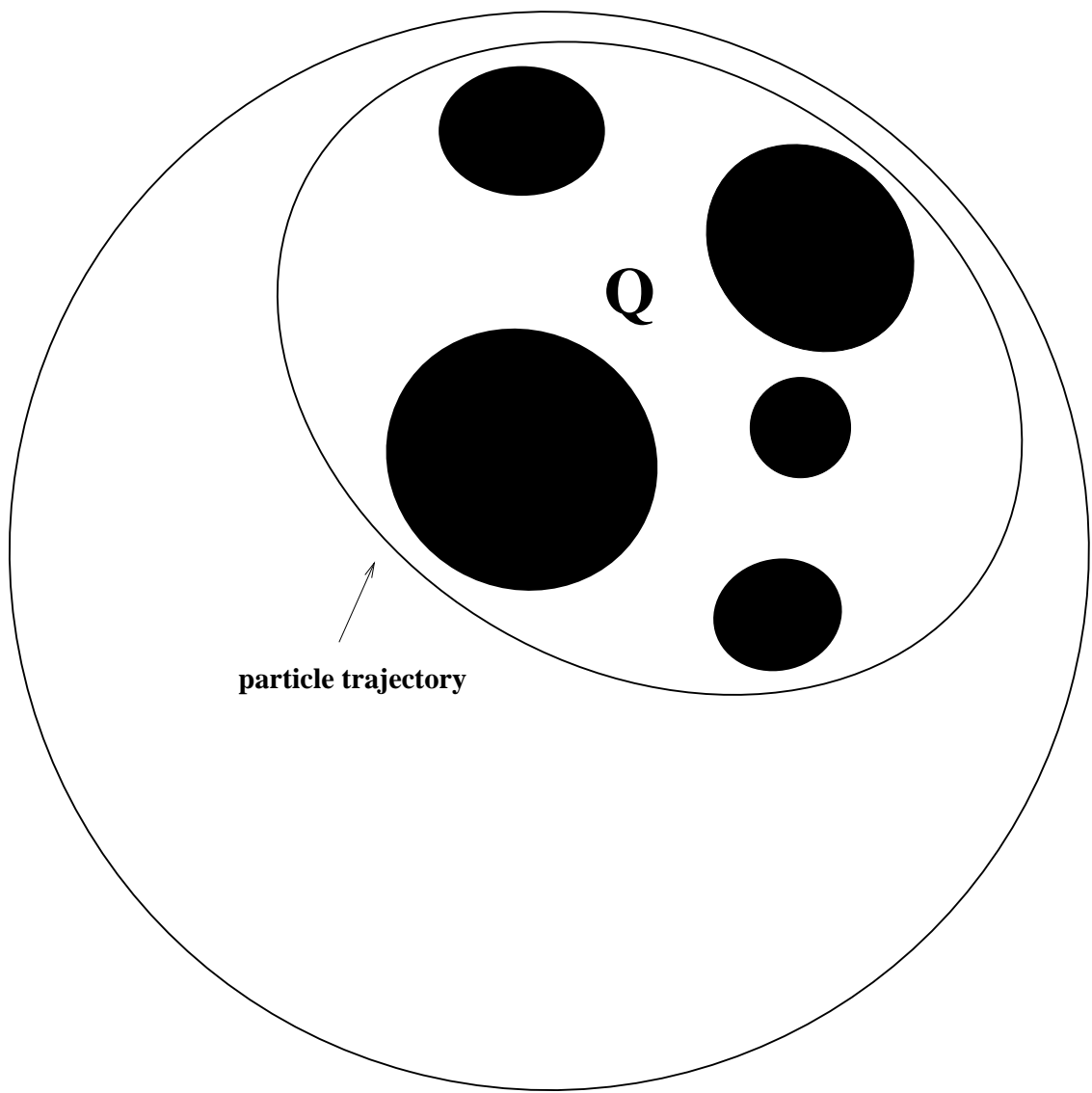
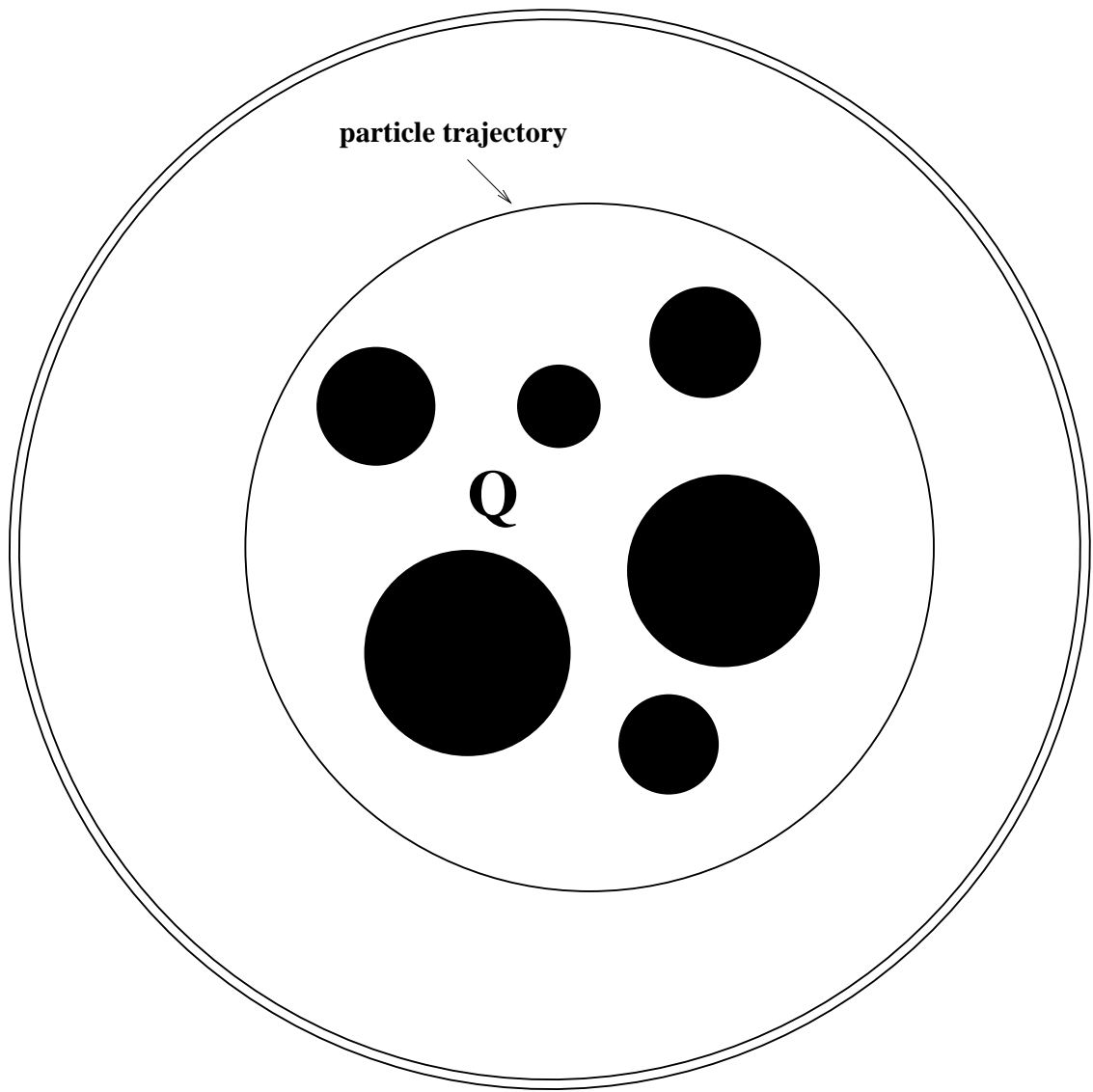
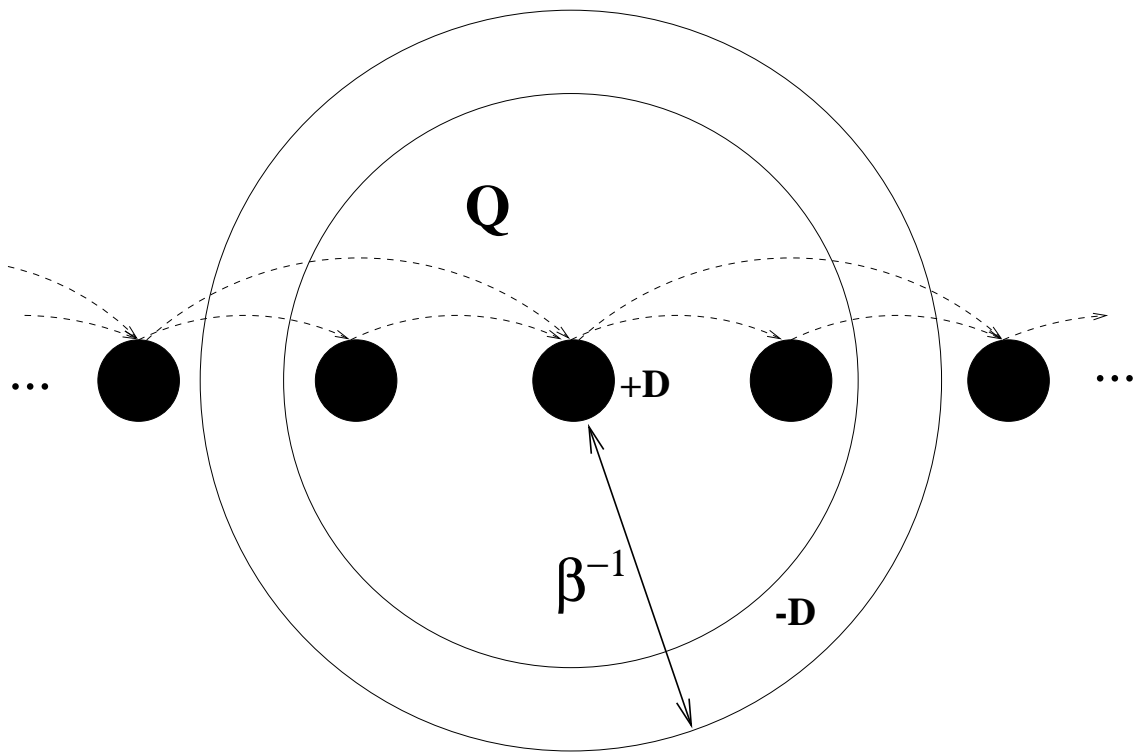
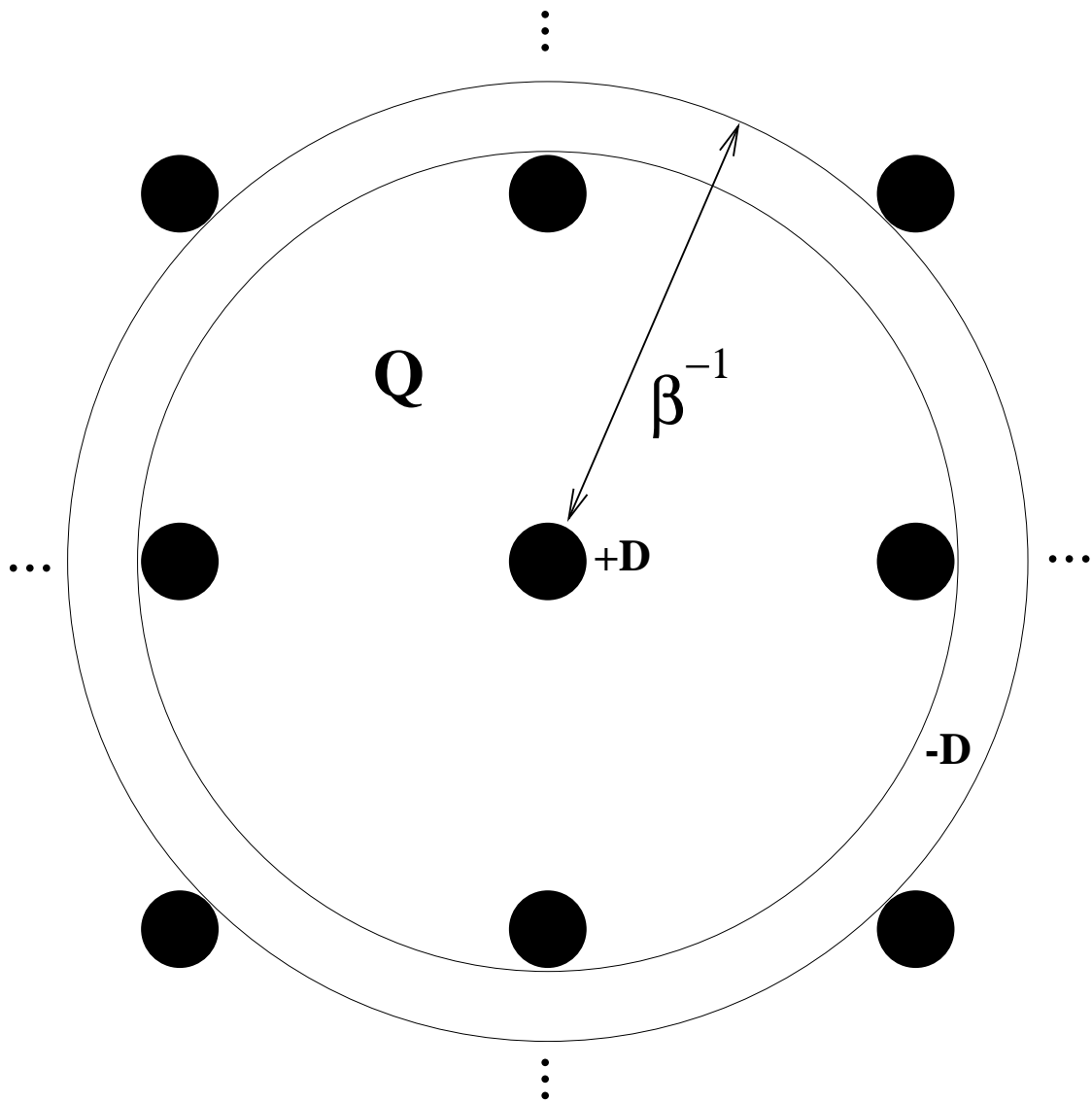
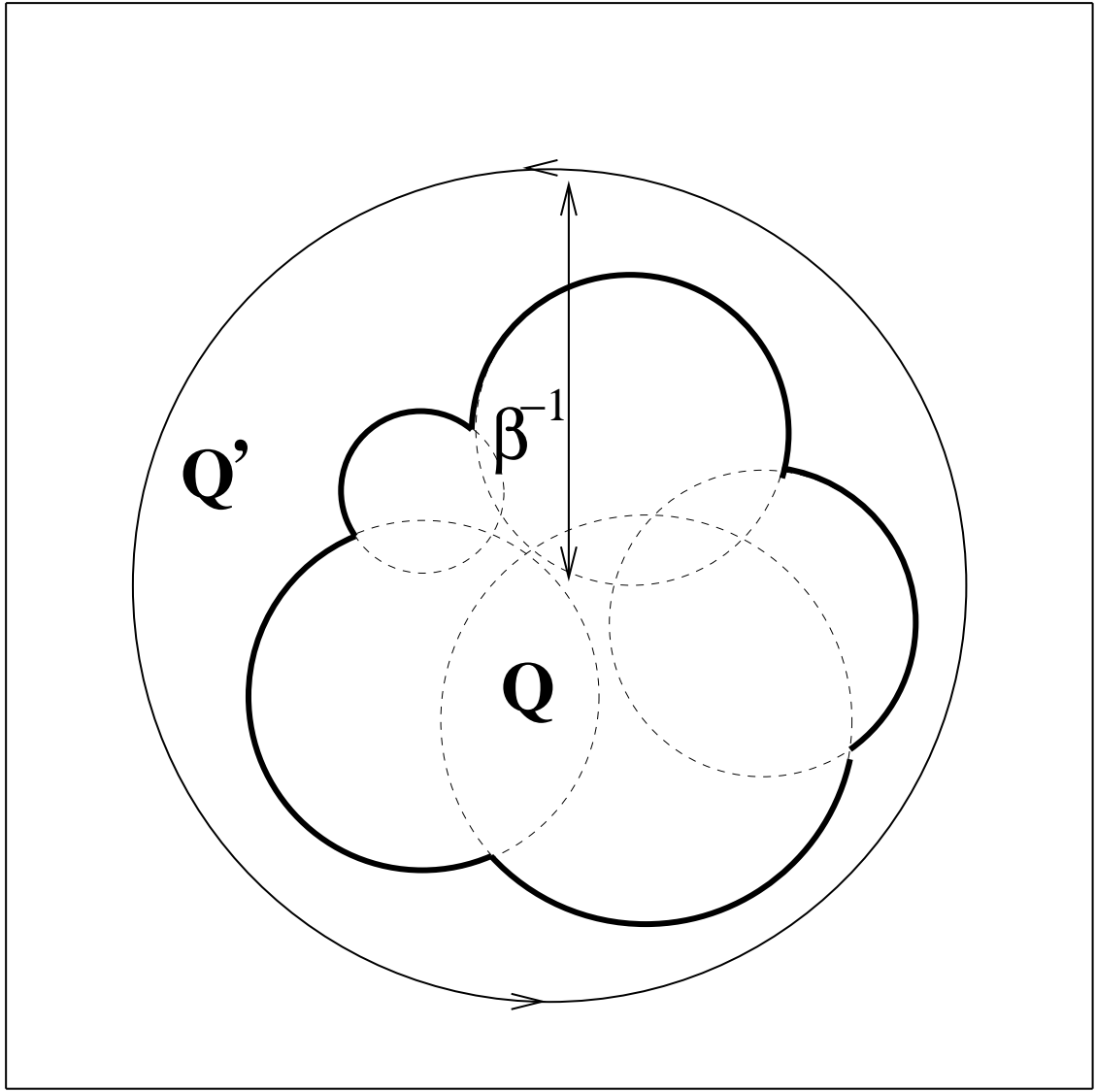
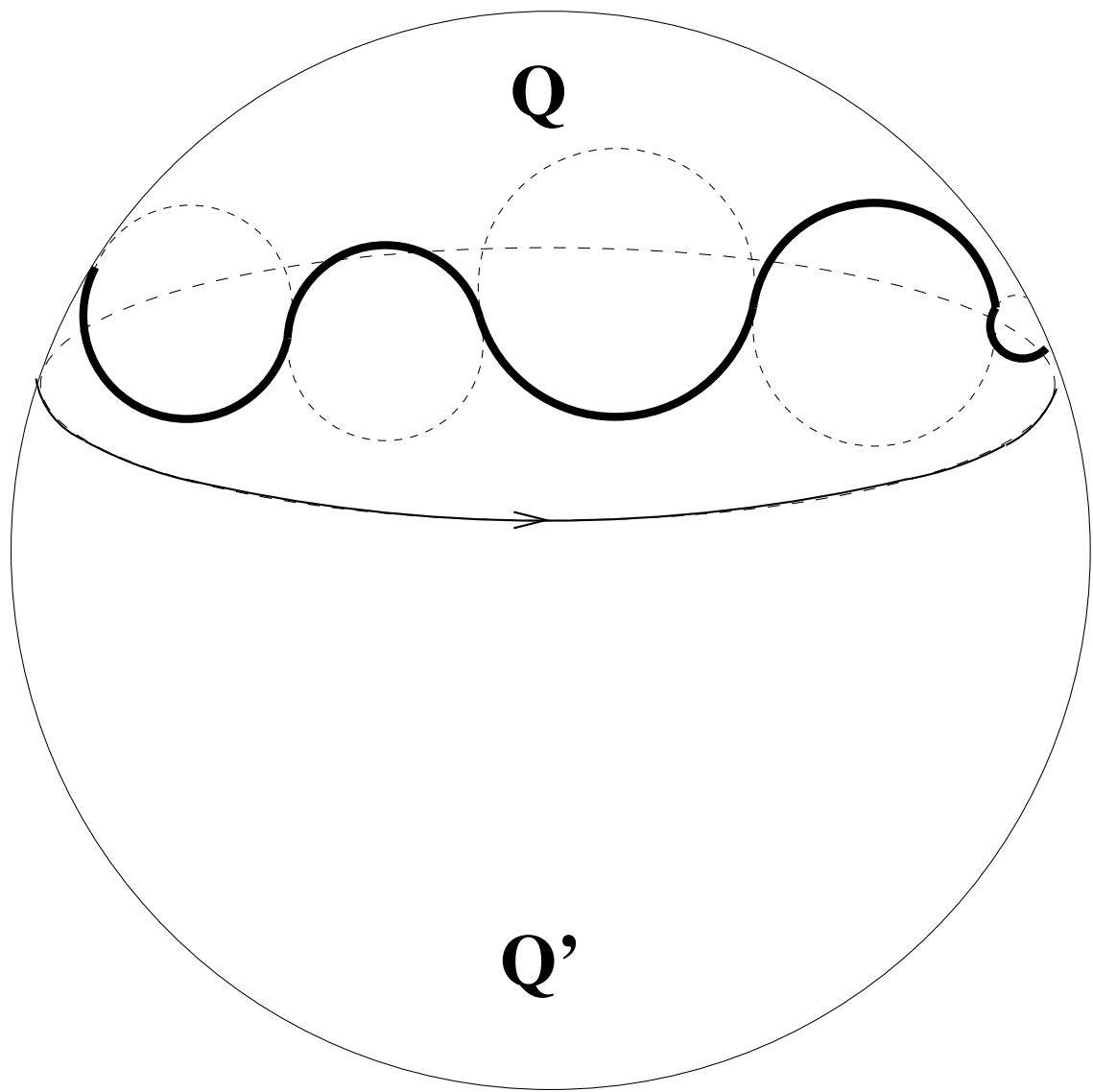
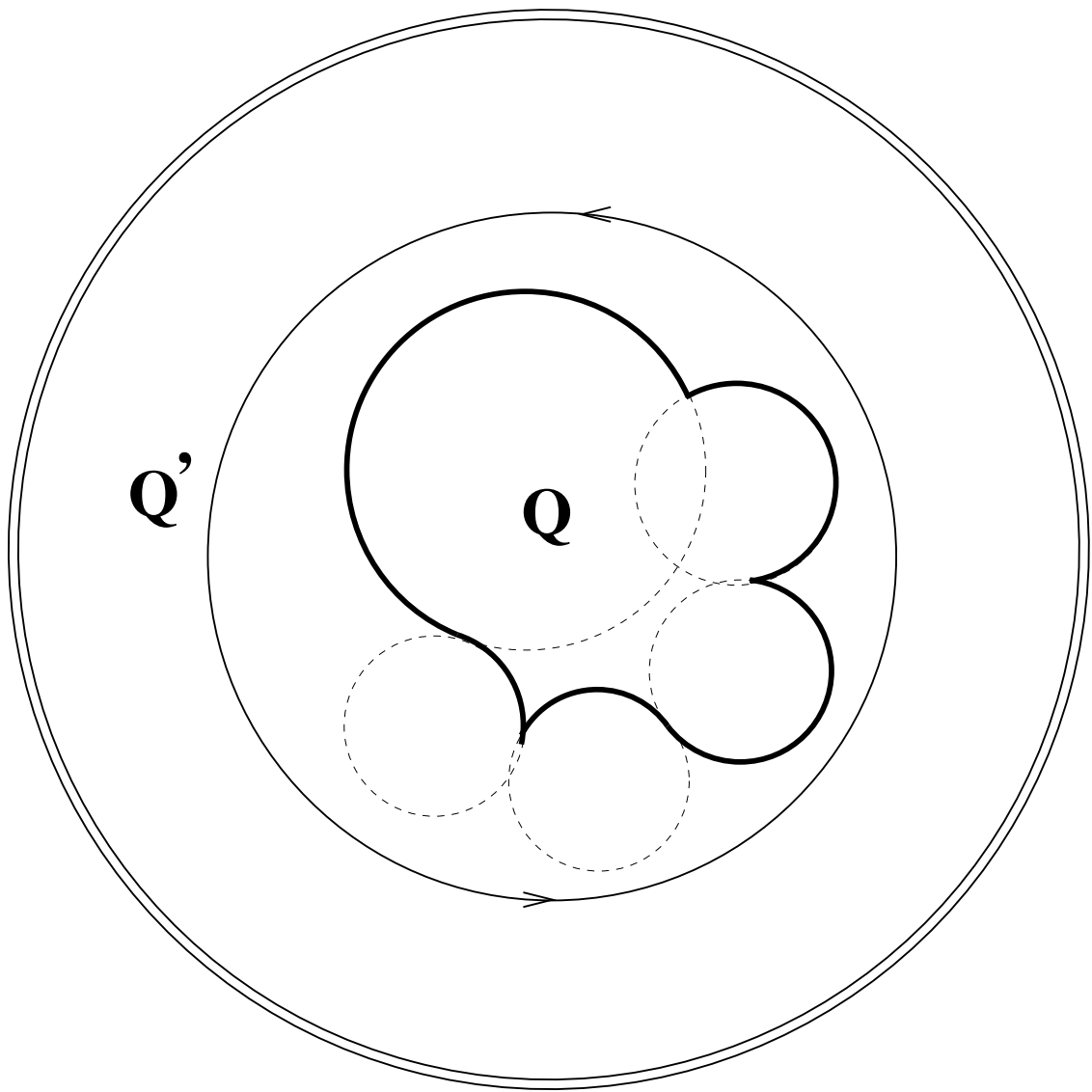
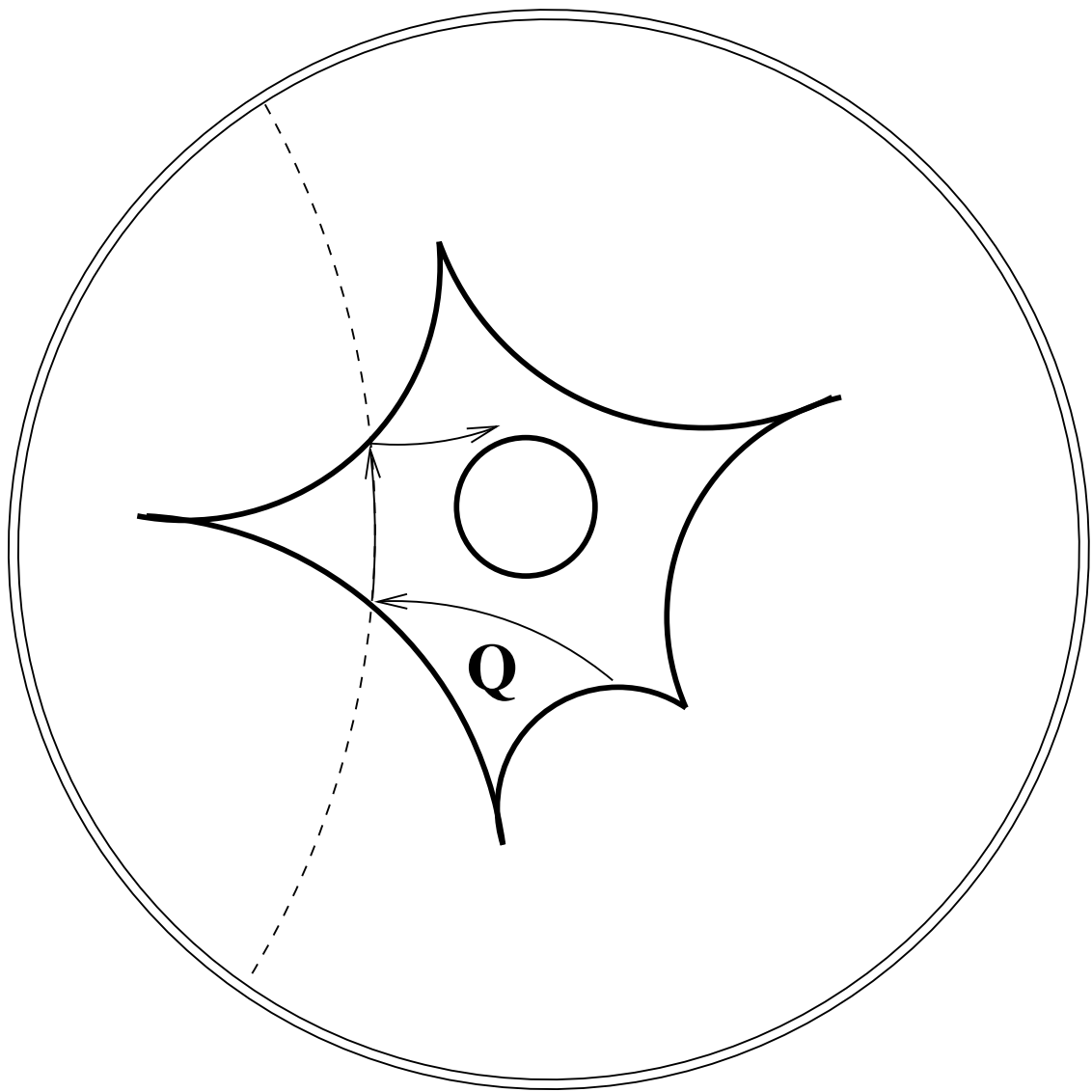
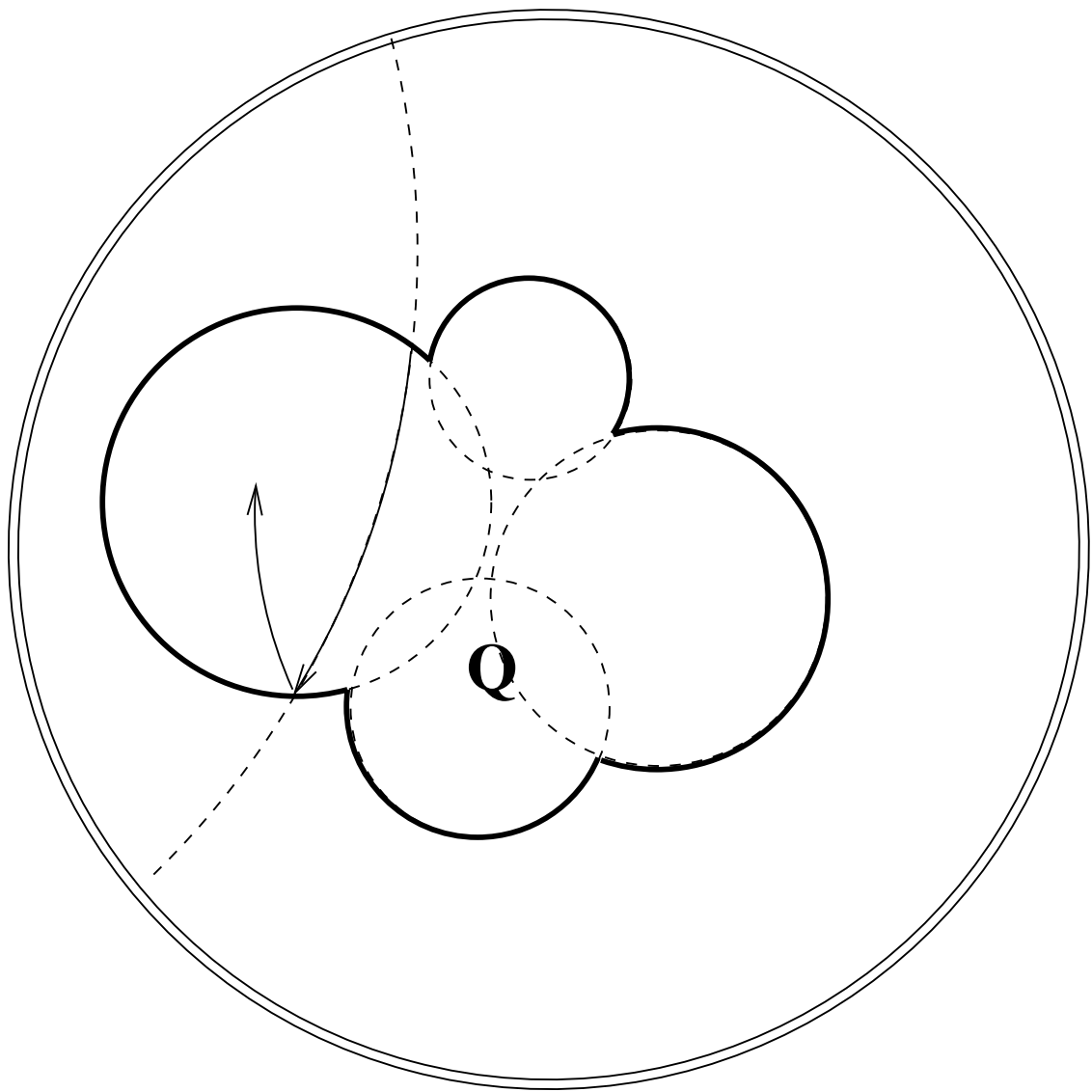
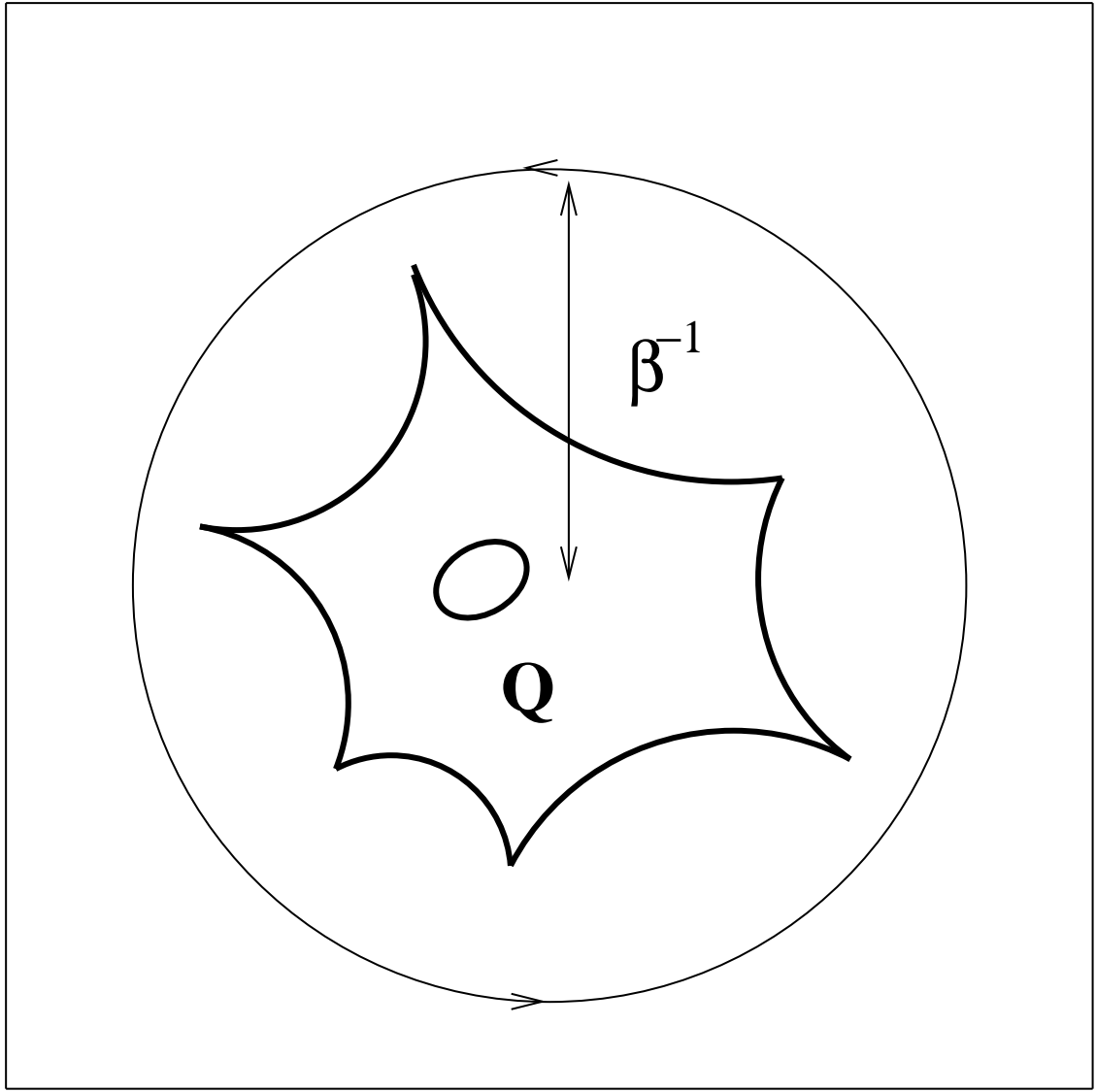
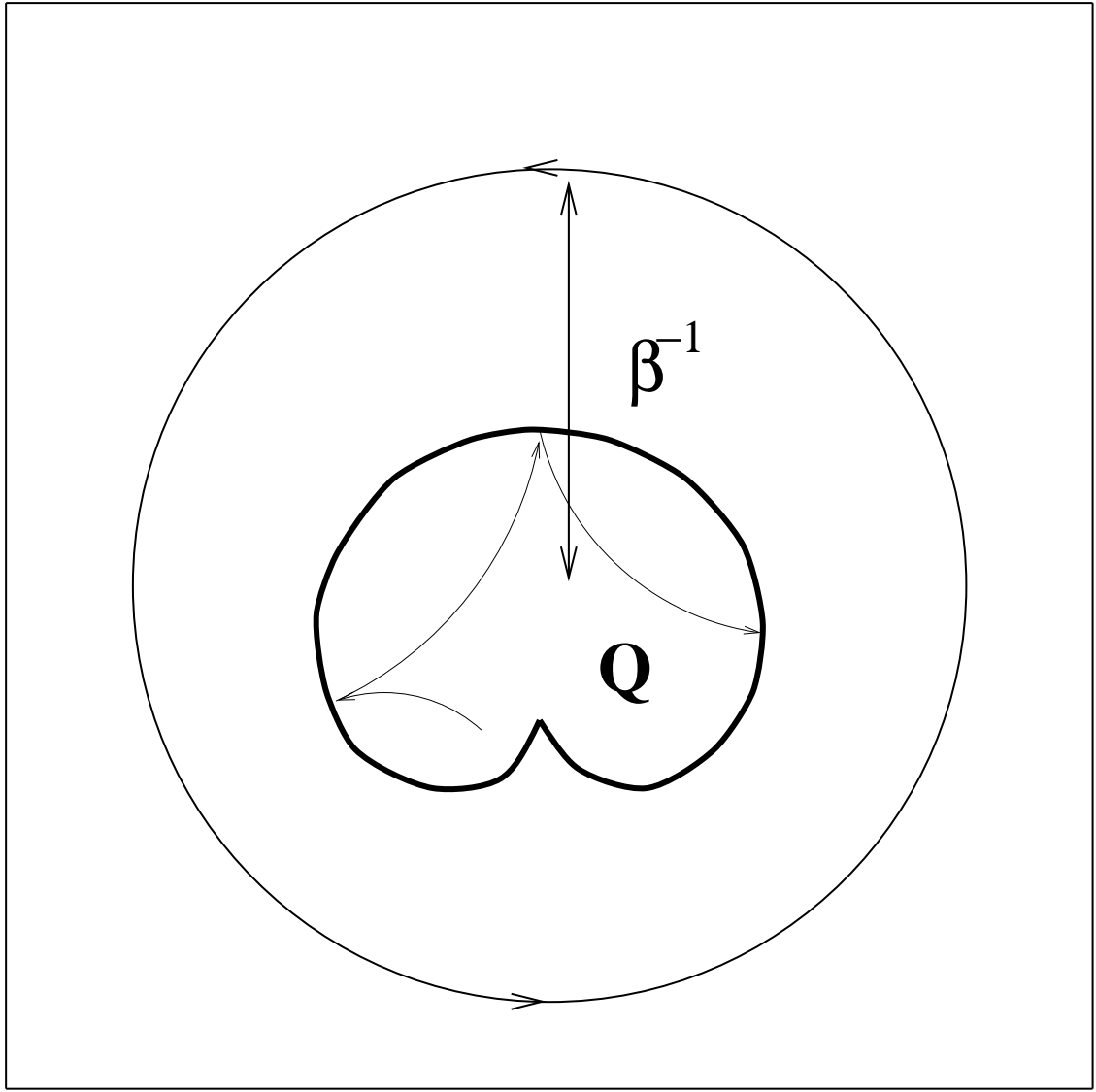
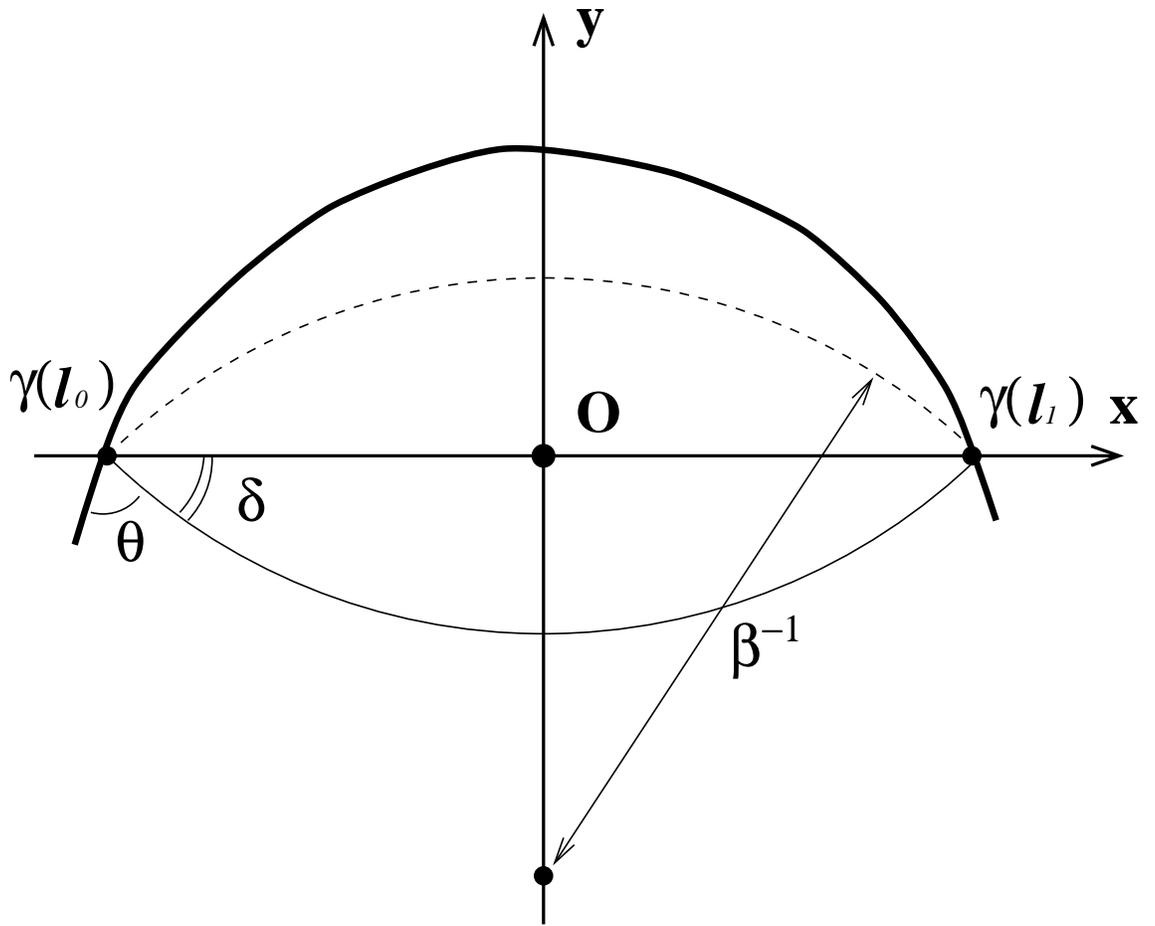
\end{document}